\documentclass{aa}  

\usepackage{graphicx}
\usepackage{txfonts}
\usepackage{hyperref}
    \hypersetup{colorlinks=true,linkcolor=blue,citecolor=blue,urlcolor=blue}
    \makeatletter
    \renewcommand*\aa@pageof{, page \thepage{} of \pageref*{LastPage}}
    \makeatother
\usepackage{algorithm} 
\usepackage{algpseudocode}
\usepackage{gensymb}

\begin{document}

   \title{Precise and efficient modeling of stellar-activity-affected solar spectra using SOAP-GPU}
   
   \titlerunning{SOAP-GPU}
   \authorrunning{Yinan Zhao et al.}

   \author{Yinan Zhao
          \inst{1}
          \and
          Xavier Dumusque\inst{1}
          \and
          Michael Cretignier\inst{2}
          \and
          Khaled Al Moulla\inst{1}
          \and
          Momo Ellwarth\inst{3}
          \and
          Ansgar Reiners\inst{3}
          \and
          Alessandro Sozzetti\inst{4}
          }
   \institute{Department of Astronomy of the University of Geneva, 51 chemin de Pegasi, 1290 Versoix, Switzerland
   \and
   Department of Physics, University of Oxford, OX13RH Oxford, UK
   \and
   Georg-August Universität Göttingen, Institut für Astrophysik und Geophysik, Friedrich-Hund-Platz 1, 37077 Göttingen, Germany
      \and
    INAF - Osservatorio Astrofisico di Torino, Via Osservatorio 20, I-10025 Pino Torinese, Italy
   \\
              \email{zhaoyinan2121@gmail.com}
             }

 
 
  \abstract
   {One of the main obstacles in exoplanet detection when using the radial velocity (RV) technique is the presence of stellar activity signal induced by magnetic regions. As the most advanced techniques to mitigate this signal are reaching a level better than one meter per second, it is difficult to evaluate their performance: instrumental systematics start to be similar in magnitude, and therefore it is impossible to know the ground truth of the stellar activity signal. In this context, a realistic simulated dataset that can provide photometry and spectroscopic outputs is needed for method development.}
   {The goal of this paper is to describe two realistic simulations of solar activity obtained from SOAP-GPU and to compare them with real data obtained from the HARPS-N solar telescope. For this purpose, both simulated spectral time series cover the time window of HARPS-N solar observation, but nothing prevents SOAP-GPU from modeling the data over different time spans.}
   {We describe two different methods of modeling solar activity using SOAP-GPU. The first models the evolution of active regions based on the spot number as a function of time. Other physical parameters are either drawn from observed solar distributions or modeled with empirical relations. The second method relies on the extraction of active regions from the Solar Dynamics Observatory (SDO) data. The location of spots and faculae on the solar disk at each timestamp are derived from the magnetogram and intensity maps and are fed into SOAP-GPU to simulate the corresponding spectra.}
   {The simulated spectral time series generated with the first method shows a long-term RV behavior similar to that seen in the HARPS-N solar observations. The effect of stellar activity induced by stellar rotation is also well modeled with prominent periodicities at the stellar rotation period and its first harmonic. The comparison between the simulated spectral time series generated using SDO images and the HARPS-N solar spectra shows that SOAP-GPU can precisely model the RV time series of the Sun to a precision better than 0.9 m/s. By studying the width and depth variations of each spectral line in the HARPS-N solar and SOAP-GPU data, we find a strong correlation between the observation and the simulation for strong spectral lines, therefore supporting the modeling of the stellar activity effect at the spectral level. The correlations are weaker for shallow lines, although it is likely that their lower signal-to-noise ratio does not allow a meaningful comparison.} 
    {We introduce two methods for modeling solar activity using SOAP-GPU. With only sunspot numbers as input, we accurately capture the long-term magnetic cycle and rotational features. Additionally, we effectively model shift and depth variations at the spectral line level by using data from SDO.  These simulated solar spectral time series serve as a useful test bed for evaluating spectral-level stellar activity mitigation techniques.}


   \maketitle
%

\section{Introduction}

The radial velocity technique was the first method used to detect a planet orbiting a solar-type star, 51 Peg b \citep{Mayor-1995Nature}, and it remains the most efficient method today for measuring planetary minimum masses. However, even considering the impressive progress that was made over the past 30 years, mainly in terms of instrumentation, this method is still limited by the stellar signal, and mainly its activity component, which prevents the detection of true Earth analogs \citep[e.g.,][]{Crass:2021aa}. In recent years, the field has moved toward probing and mitigating stellar activity in the spectral (or cross-correlation function) domain, where the stellar signal, inducing an asymmetry of spectral lines and affecting each of them differently, should be easier to disentangle from a planetary signal that simply shifts all spectral lines in wavelength \citep[e.g.,][]{Feng:2017ac,Dumusque-2018aa,Cretignier:2020aa,Collier-Cameron:2021aa,Cretignier:2021aa,AlMoulla:2022aa,Beurs:2022aa,Cretignier-2022aa,zhao:2022AJ,AlMoulla:2024aa}.

In the quest for the detection of Earth-analogs, developing stellar activity correction techniques that enable the mitigation of this perturbing signal to a few dozen cm/s requires exquisite datasets. In that direction, it is now common for ultra-stable spectrographs used for exoplanet detection during the night-time to also observe the Sun during the day. The following instruments all have solar input and provide high-resolution spectra (R$\sim$100'000) for which a radial-velocity (RV) precision better than 1 m/s is reached:  HARPS-N \citep{Dumusque-2015ApJ,Phillips:2016aa, Collier-Cameron-2019MNRAS,Dumusque-2021aa}, HARPS \citep{AlMoulla:2023aa}, EXPRES \citep{zhao:2022AJ}, NEID \citep[][]{Lin:2022aj}, KPF \citep{Rubenzahl-2023pasp}. These solar data are an ideal benchmark for developing and testing the performance of different mitigating techniques for stellar activity \citep[e.g.,][]{Collier-Cameron:2021aa,Beurs:2022aa,Langellier:2021aj,zhao:2024aa}. Although these solar observations are the closest to night-time measurements, a comparison of several solar datasets shows that at the sub-meter-per-second level, some differences are visible, likely due to instrument systematics that are not perfectly calibrated \citep[e.g.,][]{Zhao:2023ab}. In addition, even though the solar datasets now include more spectra than had been obtained on other single stars, the limited number of spectra available\footnote{The HARPS-N solar telescope, the first of its kind on sky, started operation in the summer of 2015. We now have accumulated $\sim$ 200000 spectra; however, many measurements are obtained each day. Within a day, the activity of the star which evolves on the timescale of the solar rotation, does not change significantly. Therefore it is the total number of days with observations that counts if we want to probe stellar activity. In the case of HARPS-N, this number equals approximately 1500 days.} makes it challenging to start exploring sophisticated machine learning techniques such as deep learning. For these reasons (contamination with instrumental signals and a limited amount of data), having a realistic simulation of stellar activity is crucial in order to develop and test the efficiency of mitigation techniques.

Several methods have been proposed to model solar activity to test the detection sensitivity of Earth-like planets or to derive solar activity proxies. \cite{Meunier-2010a} were likely the first to realistically model the RV of the Sun including the inhibition of convective blueshift (CB) in active regions, and demonstrated that for solar-type slow rotators this effect dominates over the flux effect induced by the contrast difference of active regions. \cite{Borgniet-2015aa} modeled solar cycle 23 in radial velocity (RV) by parameterizing the active region properties as observed on the Sun. \cite{Haywood-2016MNRAS} used the better observations provided by the Solar Dynamics Observatory (SDO), compared to the work of \cite{Meunier-2010a} that used SOHO, to model the solar RVs. They used the dopplergrams, intensity continuum, and magnetogram images from the Helioseismic and Magnetic Imager \citep[HMI/SDO,][]{Schou-2012SoPh} to derive the RV components including the CB inhibition and the flux imbalance seen in facula and spot regions. All these simulations are providing realistic RVs affected by stellar activity for the Sun, however, without modeling properly the solar spectra. The first such an attempt was likely the work performed by \cite{Gilbertson:2020aa} that used a modified version of SOAP 2.0 \citep{Dumusque-2014b}, but the simulation is likely too simplistic as it only parameterizes the spot region configurations without considering faculae, which are the main contributors of stellar signal for the Sun. In addition, SOAP 2.0 is rather limited by computational performance in the number of spectra that can be modeled.

Recently, the SOAP-GPU code was developed to lift the SOAP 2.0 constraint on computational effort thanks to GPU performance \citep{Zhao-2023A&A}. In addition to being more efficient than SOAP 2.0, SOAP-GPU also includes more physics based on recent solar observations obtained by \cite{Lohner-Bottcher:2019aa} and \cite{Ellwarth-2023aa}. The simulation now includes realistic center-to-limb variation (CLV) of the quiet Sun spectrum. Such observations at different limb-angles, which cover the full visible spectrum, still do not exist for active regions. Therefore SOAP-GPU takes as input the CLV observed in \cite{Cavallini-1985a} for the FeI line at $6301.5008\AA$, and propagates the observed bisector to all spectral lines.
Thanks to SOAP-GPU, it is now possible to model the effects of solar activity at the spectral level with likely the best precision possible.

In this paper we present two simulated solar spectral time series datasets modeled with SOAP-GPU. The first, described in Sect.~\ref{sec2}, uses as input the solar spot number time series to model solar activity based on the known properties of solar active regions. The second, described in Sect.~\ref{sec3}, uses as input the SDO images to extract at each time the location and size of active regions. As this is as close as possible to reality, we then compare the modeled spectra with the ones derived from the HARPS-N solar telescope. Finally, we discuss and draw our conclusions in Sect.~\ref{sec4}. The simulated datasets and the simulation code are publicly available on Zenodo and Github. \footnote{Data available at \url{https://doi.org/10.5281/zenodo.14262853} and code available at \url{https://github.com/YinanZhao21/SOAP_GPU}.}

\section{Solar spectral time series modeling using spot number}\label{sec2}

Two main factors are taken into account when using SOAP-GPU to simulate activity-affected spectral time series, i) the properties of active regions and ii) the properties of input spectra for both the quiet and active regions. The properties of active regions, both in the spatial and the temporal domain, are well described in \cite{Borgniet-2015aa} and we present in Sect.~\ref{sec2.1} the parametrization we used as input of SOAP-GPU to model solar activity. In Sect.~\ref{sec2.2} we describe the input spectra that are used in SOAP-GPU to model the quiet and active regions. Finally, the simulated spectral time series is presented in Sect.~\ref{sec2.3}. 

Throughout this paper, the differential rotation of the solar disk is included according to the equation $\omega = \omega_{0}+\omega_{1}\sin^{2}(\theta)$, where $\omega_{0} = 14.371^{\circ}/\rm{day}$ and $\omega_{1} = -2.587^{\circ}/\rm{day}$ \citep{Borgniet-2015aa, Zhao-2023A&A}. We assume that both quiet and active regions have constant temperatures, which are set to $\rm{T_{eff}} = 5778K$ for the quiet photosphere and to $\rm{T_{eff}} = 6028K$ and $\rm{T_{eff}} = 5115K$ for the facula and spot regions, respectively \citep{Meunier-2010a}.

  \begin{figure*}[htbp]
  \includegraphics[scale=0.45]{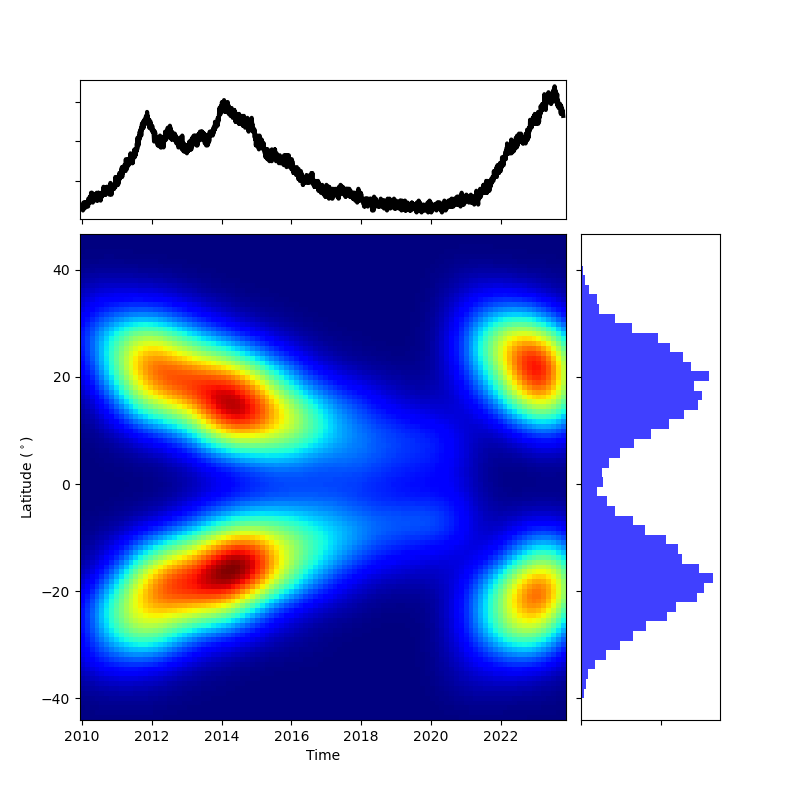}
  \includegraphics[scale=0.45]{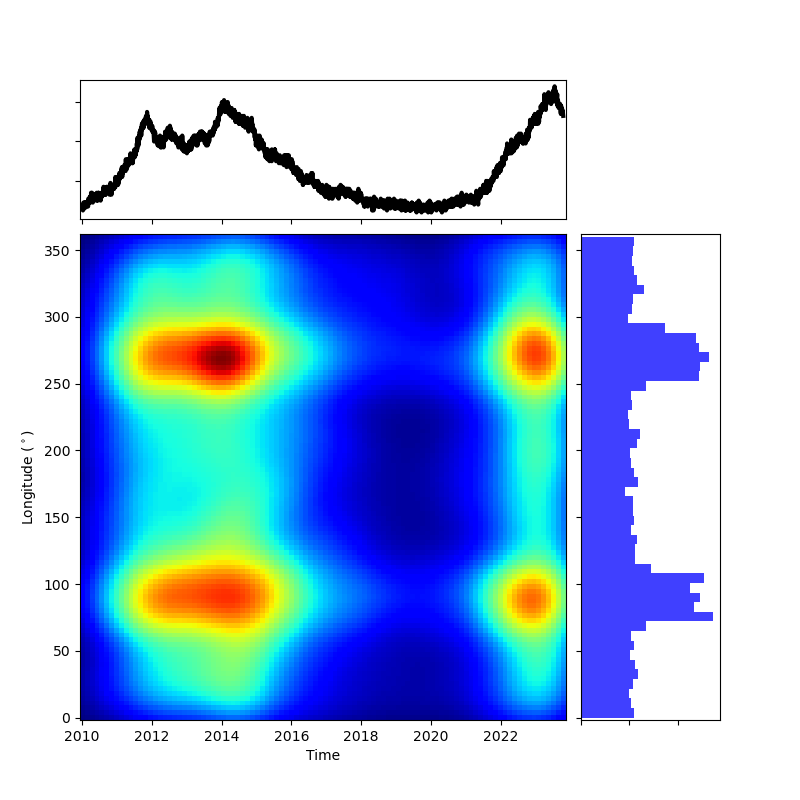}
  \caption{Location of active regions initialized with spot number evolution curve as a function of time. \emph{Left}: Latitude evolution of injected active regions. The spot number evolution curve derived from the Solar Influences Analysis Data Center (SIDC) is shown in the top panel. The latitude distribution shown in the histogram on the right indicates that the north-south asymmetry factor is 0.5. The butterfly diagram is shown in the density map. \emph{Right:} Longitude evolution of injected active regions. The spot number evolution curve derived from the SIDC is shown in the top panel. The longitude distribution shown in the histogram on the right indicates that the two considered active longitude regions are separated by 180 degrees.}
  \label{Fig_lat_long}%
  \end{figure*}

\subsection{Active region properties} \label{sec2.1}

The number of spots on the solar surface is important in order to capture the short-term solar activity at the rotational and half of the solar rotational periods, and also the long-term activity induced by the solar magnetic cycle. For this study we used the spot number time series provided by the Solar Influences Data Analysis Center (SIDC) \citep{sidc}. \footnote{Data available at \url{https://www.sidc.be/SILSO/home}.} The spot number $n$ is described by the Wolf number $R$, which is defined by $R = s + 10g$, where $s$ is the number of isolated spots and $g$ is the number of complex spot groups. The numbers of these two spot types are obtained by using the fraction of isolated spots number versus all the spot numbers. In this simulation we used a value of 0.4, as mentioned in \citep{Borgniet-2015aa}. So the numbers of isolated spots and complex spot groups are $s = 0.4R$ and $g = 0.06R$, respectively. Another important aspect on the timescale of the solar magnetic cycle is the migration of active regions from high latitudes at the beginning of the cycle toward the equator at the end. We used the empirical equation in \cite{Hathaway-2011SoPh} to describe this butterfly diagram:
\begin{equation}\label{eq:1}
    \bar{\lambda }(t) = 28^{\circ} \exp{[-(t - t_{0})/90]}.
\end{equation}
Here $\bar{\lambda }$ is the average latitude of the active regions. $t$ is time in units of months, and $t_0$ is the cycle starting time. A magnetic cycle of 132 months is used to repeat the butterfly diagram. A dispersion of $6^{\circ}$ is added to the butterfly diagram to model realistic observations. As described in \cite{Borgniet-2015aa}, the active regions are not randomly distributed along the longitude of the solar disk. Two persistent active regions along the longitude are separated by $180^{\circ}$. The active longitudes are also included in our SOAP-GPU simulation. In \cite{Borgniet-2015aa}, the active longitude was formed by injecting a fraction of new sunspots into a confined area where existing sunspots were already present. For this paper we followed the same method. We defined two active longitude regions with average longitudes of $90^{\circ}$ and $270^{\circ}$ with a dispersion $20^{\circ}$. Forty percent of the spots are generated in the active regions, while the rest of the spots are uniformly generated along the longitude. The latitude distribution and longitude distribution of the active regions initialized by the 14 years of spot number data are illustrated in Fig.~\ref{Fig_lat_long}. Since we adopt the value of 11 years as the average duration of the magnetic cycle, only one butterfly diagram is fully shown in the latitude distribution. For simplicity, we chose to keep the active longitude regions fixed between cycles as an analysis by \cite{Berdyugina:2003aa} suggests that the same two active longitudes, separated by 180 degrees, have persisted on the Sun for over 120 years. While this may not hold for other stars, we also explored varying these longitudes between the two modeled cycles. As expected, the difference in the long-term is negligible as the inhibition of convective blueshift, related to the active region filling factor rather than its position on the disk, dominates the RV signal over magnetic cycles. The evolution of the individual active regions is also taken into account in the SOAP-GPU simulation. As \cite{Borgniet-2015aa} pointed out, the timescale of the growing phase of active region is much smaller than the timescale of the solar rotation; therefore, the growing phase is unlikely to contribute much to the solar activity on the timescale of the solar rotation. In the simulation, we thus only take into account the decay phase of active regions. The initial size of spot regions used in the simulation follows a log-normal distribution \citep{Baumann-2005aa,Borgniet-2015aa,Gilbertson:2020aa},
\begin{equation}\label{eq:2}
    \frac{dN}{dA} = \frac{1}{\sigma A \sqrt{2\pi}} \exp{[-\frac{(log(A) - \mu)^2}{2 \sigma^2}]},
\end{equation}
where $\sigma = log(\sigma_{s})$ and $\mu = log(\Bar{A}_{s})$, $\Bar{A}_{s}$ and $\sigma_{s}$ are the mean initial size, and the corresponding standard deviation is in units of micro hemispheres ($\mu$Hem). In the setup of the simulation we followed the same initialization as used in \cite{Borgniet-2015aa}: isolated spots have $\Bar{A}_{s} = 46.51$ and $\sigma_{s} = 2.14$ and complex spot groups have $\Bar{A}_{s} = 90.24$ and $\sigma_{s} = 2.49$. We assumed that active regions follow a linear decay law. The average decay law also follows a log-normal distribution \citep{Martinez-1993aa,Borgniet-2015aa,Gilbertson:2020aa},
\begin{equation}\label{eq:3}
    \frac{dN}{dD} = \frac{1}{\sigma_{logD} D \sqrt{2\pi}} \exp{[-\frac{(log(D) - \mu_{logD})^2}{2 \sigma_{logD}^2}]},
\end{equation}
 where $\sigma_{logD} = log(\sigma_{D})$ and $\mu_{logD} = log(\Bar{D})$, $\Bar{D}$ and $\sigma_{D}$ are the median decay rate, and the corresponding standard deviation is in units of $\mu$Hem/day. We used $\Bar{D} = 14.8, 30.9$ and $\sigma_{D} = 2.01, 2.14$ for the isolated spots and complex spots group, respectively. For the simulation of facula regions, we assumed that every spot region is surrounded by a facula region. The initial size of the facula region is governed by the spot-to-facula ratio. We used a uniform distribution with a boundary from 5 to 10 to model the spot-to-facula ratio. The decay rate of faculae also follows Eq.~\ref{eq:3} with $\Bar{D} = 20.0 $ and $\sigma_{D} = 0.77$ \citep{Borgniet-2015aa}. For every timestamp, the number of spots was obtained by using the corresponding spot decay rates. If the number of spots was smaller than the expected number derived from the SIDC data, we injected new spots. This method may lead to instances where the actual injected number exceeds the expected number. To quantify this effect we conducted a Monte Carlo simulation, and found that for spot numbers less than 20 there is a $20\%$ excess injection, while for spot numbers greater than 20, the excess injection is only $2\%$. We agree that the relatively simple choices made in this section regarding active longitude properties, active region sizes, and their evolution can be discussed further in depth. However, the goal here is not to provide simulations tailored to match any specific star, but rather to demonstrate that with simple statistical assumptions it is possible to model realistic RVs for solar-type stars, including rotational and long-term magnetic effects, which is demonstrated in Fig.~\ref{Fig_RV_periodogram}.

\subsection{Input spectra properties} \label{sec2.2}

SOAP-GPU models, for each point on the visible solar hemisphere, the effect of stellar rotation, differential rotation, and limb-darkening. In addition, by injecting into SOAP-GPU different spectra as a function of $\mu=\cos(\theta)$, where $\theta$ is the center-to-limb angle, it is also possible to model center-to-limb variations, as shown in \cite{Zhao-2023A&A}.

In practice, SOAP-GPU needs as input three spectral cubes for the quiet photosphere for faculae and for spots. As SOAP-GPU is able to model the Sun, and other stars as well, the code can use high-resolution stellar spectra from the PHOENIX database \citep{Husser-2013aa} as input. These spectra however do not include realistic convection and therefore spectral lines do not have a realistic shape, which is crucial to properly modeling spectra affected by stellar activity. As explained in more detail in \cite{Zhao-2023A&A}, we can use the quiet sun observations at different $\mu$ angles from \citet{Lohner-Bottcher:2019aa} to modify the bisector of PHOENIX spectra at different disk positions, and therefore obtain the desired spectral cube for quiet photospheric regions. Recently, \cite{Ellwarth-2023aa} published the IAG solar atlas that includes high-resolution spectra of the quiet Sun taken at 14 heliocentric positions for the quiet solar disk, from $\mu=1$ (disk center) to $\mu=0.2$ (limb). It is therefore also possible to directly inject those spectra in the spectral cube for the quiet Sun. We differentiate between the two options, and refer to them as the PHOENIX and IAG cases.

\begin{figure}[htbp]
  \includegraphics[scale=0.25]{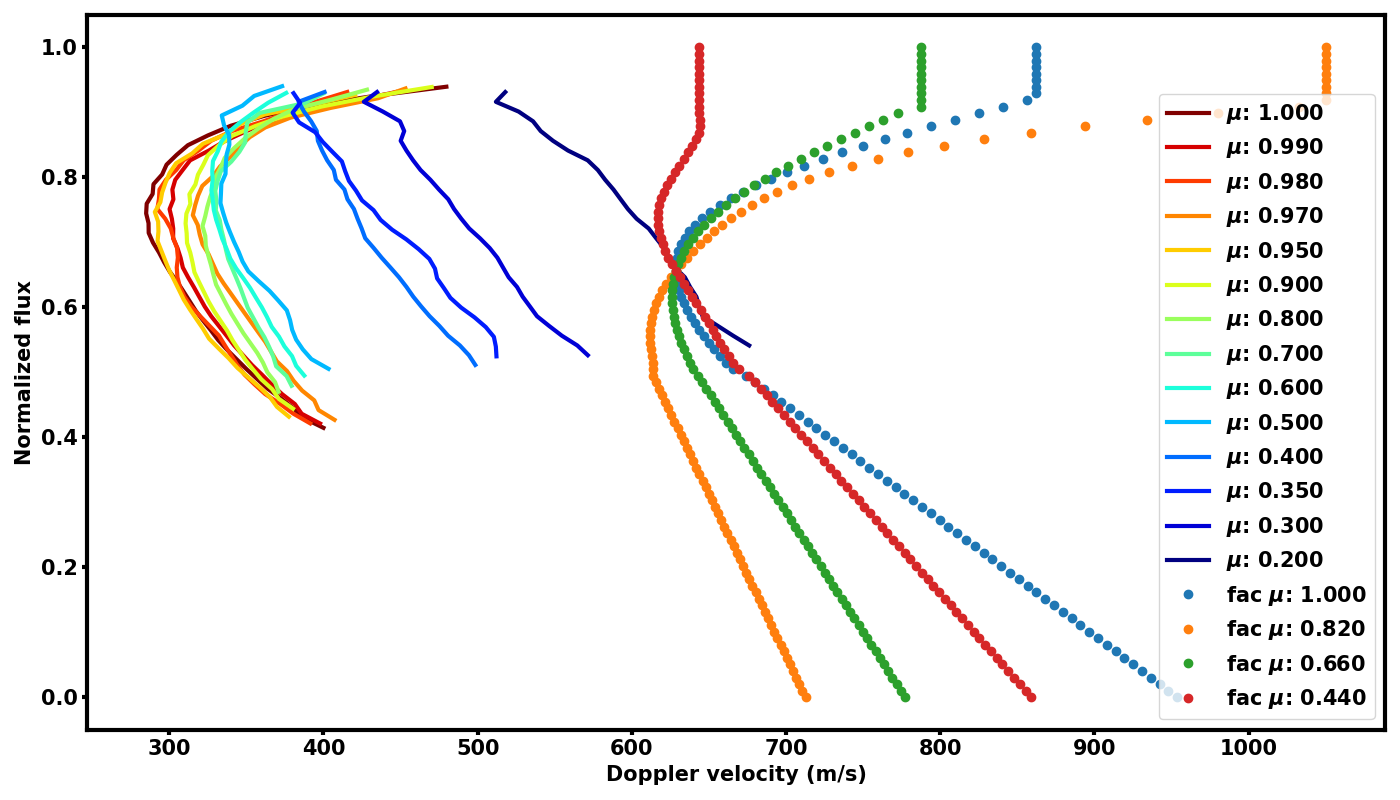}
  \caption{Line bisectors of quiet and active solar regions from the disk center ($\mu = 1.0$) to the limb ($\mu = 0.2$). \emph{Continuous lines:} Line bisectors of quiet solar disk at 14 different heliocentric positions measured from \cite{Ellwarth-2023aa}, similar to Fig. 11 of \cite{Ellwarth-2023aa}. Unlike in Fig. 8 of \cite{Zhao-2023A&A}, we do not inject the bisectors of the quiet region into the spectra, as they are naturally included. \emph{Dotted lines:} Fit of the bisectors of the FeI $6301.5008\AA$ spectral line inside a facula region, as measured by the Fabry-Perot interferometer at the Donati Solar Tower \citep[][]{Cavallini-1985a}. Below a depth of 0.5, a linear fit is performed, while a fifth-order polynomial is used to model the top part of the bisector. To prevent unrealistic values when interpolating the polynomial above a normalized flux of 0.9 where no measurement exists, we selected the most redshifted part of the top bisector, explaining the vertical values for very shallow depths. The bisector of active regions at different $\mu$ angles are all shifted by 350 m/s based on our hypothesis that convection is fully suppressed in magnetic regions.}
  \label{Fig_BIS}%
  \end{figure}

For active regions, we used the results of \citet{Cavallini-1985a}, who observed the bisector variation of the FeI line at $6301.5008\AA$ as a function of $\mu$. This is the only observation available in the literature. Although using bisector changes from a single line may not perfectly model the CB effect since the convective blueshift inhibition may affect different lines in varying ways, this approximation should be valid to first order as all lines are redshfited and warped by the CB effect. As shown in Figs. 5 through 8 in \cite{Gray-2009}, the selected lines in that paper generally follow common bisectors, with some departure due to line blending. We note, however, that the results from \cite{Gray-2009} should be interpreted with caution as the analysis by \citet{Palumbo:2024aa} of observed spectra from the LARS spectrograph indicates that line bisectors of other lines \citep[not presented in][]{Gray-2009} differs quite significantly from the general trend, mainly for the very strong lines. Before injecting the appropriate bisectors as a function of $\mu$, the original bisectors in the input spectra are removed by fitting line bisectors with polynomial functions. A detailed description of bisector injection can be found in Section 4.3.2 of \cite{Zhao-2023A&A}. In the case of a spot, a PHOENIX spectrum with $\rm{T_{eff}} = 5115K$ is used and, as in the case of the quiet photosphere, the bisector of spectral lines is changed to produce in the end the spectral cube for the spots. The spectral cube for the faculae is generated in the same way, just starting from a PHOENIX spectrum with $\rm{T_{eff}} = 6028K$. We note that for both active regions, we have to adjust the velocity offset between quiet and active regions to properly take into account the inhibition of CB. This is done by assuming that the inhibition of convection in active region is the same as the maximum inhibition that happens in quiet regions between the center of the disk and the very limb (in our case at $\mu=0.2$). We can see in Fig.~\ref{Fig_BIS} the quiet bisectors derived from \cite{Ellwarth-2023aa} and the active bisectors from \citet{Cavallini-1985a} shifted to match the maximum CB of the quiet spectra.\footnote{When the quiet Sun spectral cube is derived in the PHOENIX case, the quiet Sun bisectors are slightly different from the ones derived from the IAG case, and therefore the CB inhibition of the active region bisectors will be slightly different.}

\subsection{Spectral time series results} \label{sec2.3}

\begin{figure*}[htbp]
  \centering    
  \includegraphics[scale=0.23]{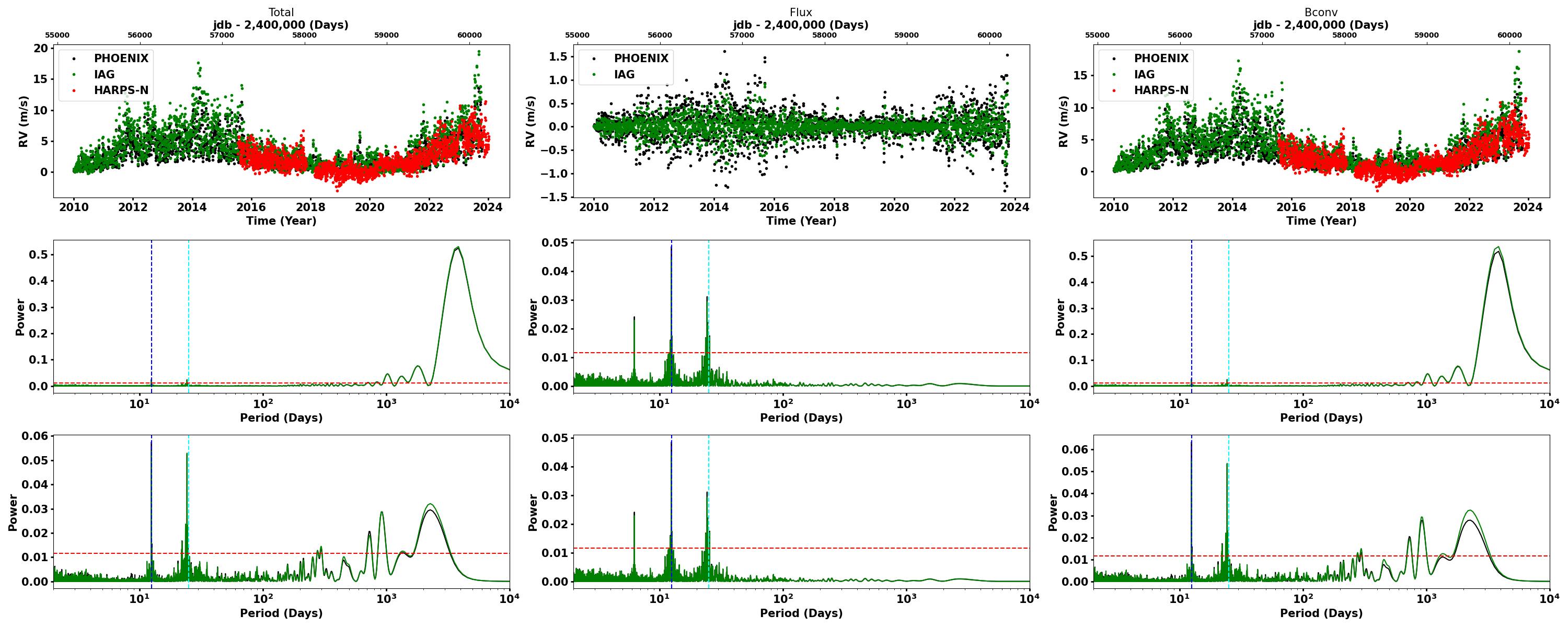}
  \caption{RV time series and periodograms from the solar simulations using spot number as input. In each panel we show the results for the PHOENIX and IAG cases in black and green, respectively. \emph{Top:} From left to right, we show the RVs of the total effect, flux effect, and the effect of CB inhibition. The RVs of the HARPS-N solar observation, in red, are also illustrated in the left and right panels for comparison.
  Middle: Corresponding periodograms for each effect. The FAP level of $0.1\%$ is represented by the red dashed line. The half rotation and the rotation period of the Sun are indicated with blue and cyan dashed lines.
  Bottom: Corresponding periodograms of each effect, but with the magnetic cycle removed. Since the RVs from the flux effect are not affected by the long-term trend induced by the magnetic cycle, the periodogram remains the same.}
  \label{Fig_RV_periodogram}%
  \end{figure*}

Using SOAP-GPU with the inputs described above, we simulated one solar spectrum every two days for 14 years, between 2010 and 2024, to cover solar magnetic cycle number 24 and the HARPS-N solar observations. Based on an active region map that is generated at each timestamp from the spot number, SOAP-GPU integrates over all active regions at the given time to generate an integrated spectrum, which corresponds to the difference between the quiet and active Sun at the position of active regions. This integrated spectrum is then subtracted from the integrated spectrum of the full disk covered with quiet regions only, therefore producing the full disk integrated spectrum affected by activity. In total, our simulation represents 2555 $\times$ 3 spectra as SOAP-GPU gives as output the solar spectrum affected by all activity components, but also affected either by the inhibition of CB or by the flux effect. The RV values are derived by cross-correlating the obtained spectra with the ESPRESSO G2 mask. The RV time series of different stellar activity effects and the corresponding periodograms are shown in Fig.~\ref{Fig_RV_periodogram}.

Based on the simulation using only the spot number, the stellar activity affected RVs are dominated by the inhibition of CB, as presented in Fig.~\ref{Fig_RV_periodogram}. Overall, the difference between the PHOENIX and IAG cases is very minor. This is not surprising since both simulations used the same set of active region maps generated from the spot number. Since the simulation covers the HARPS-N solar observation, we also compared the simulated RVs with the HARPS-N solar RVs. Although the real location and size of active regions for a given epoch cannot be modeled by only using the spot number and empirical equations, the long-term magnetic cycle presented in the HARPS-N solar data can be well modeled by the simulations, as shown in the top left and top right panels of Fig.~\ref{Fig_RV_periodogram}. The periodograms of the total effect and the CB effect are dominated by a long-term trend associated with the magnetic cycle, implying that in the long-term the CB effect dominates over the flux effect. The periodogram of the flux effect is itself dominated by stellar activity at the rotational timescale and its respective harmonics (one-half and one-third of the solar rotation). Once the long-term trend is removed by fitting a third-order polynomial on the RVs of the total and CB effects, stellar activity at the rotational period and the respective harmonics stands out, as in the flux effect case.

\section{Solar spectra modeling using SDO data}\label{sec3}

As SOAP-GPU requires as input an active region map at each timestamp (see Sect.~\ref{sec2.3}), we can inject for each time of the simulation the observed location and size of the solar active regions (spots and faculae) as extracted from the different images produced by SDO. Compared with the simulation that uses only the spot number time series as input (see Sect.~\ref{sec2}), our simulation using SDO image information as input does not require additional assumptions on the initial conditions of the active regions, active region evolution curves, and parameterization of latitude and longitude distributions since all those information are included in the SDO images. By using SDO data, SOAP-GPU is able to model the activity-affected spectra more precisely than the simulation generated with only the spot number time series, which allows us to then compare in detail the simulated spectral time series with the HARPS-N solar spectra. In Sect~\ref{sec3.1}, we describe the different processing steps of SDO images to derive the active region map used as input to SOAP-GPU. The results of the simulation are present in Sect~\ref{sec3.2}. We compare the SDO simulation with HARPS-N solar spectra at the spectral level in Sect~\ref{sec3.3}.

\subsection{SDO data preprocessing} \label{sec3.1}

\begin{figure*}[htbp]
   \centering
   \includegraphics[scale=0.55]{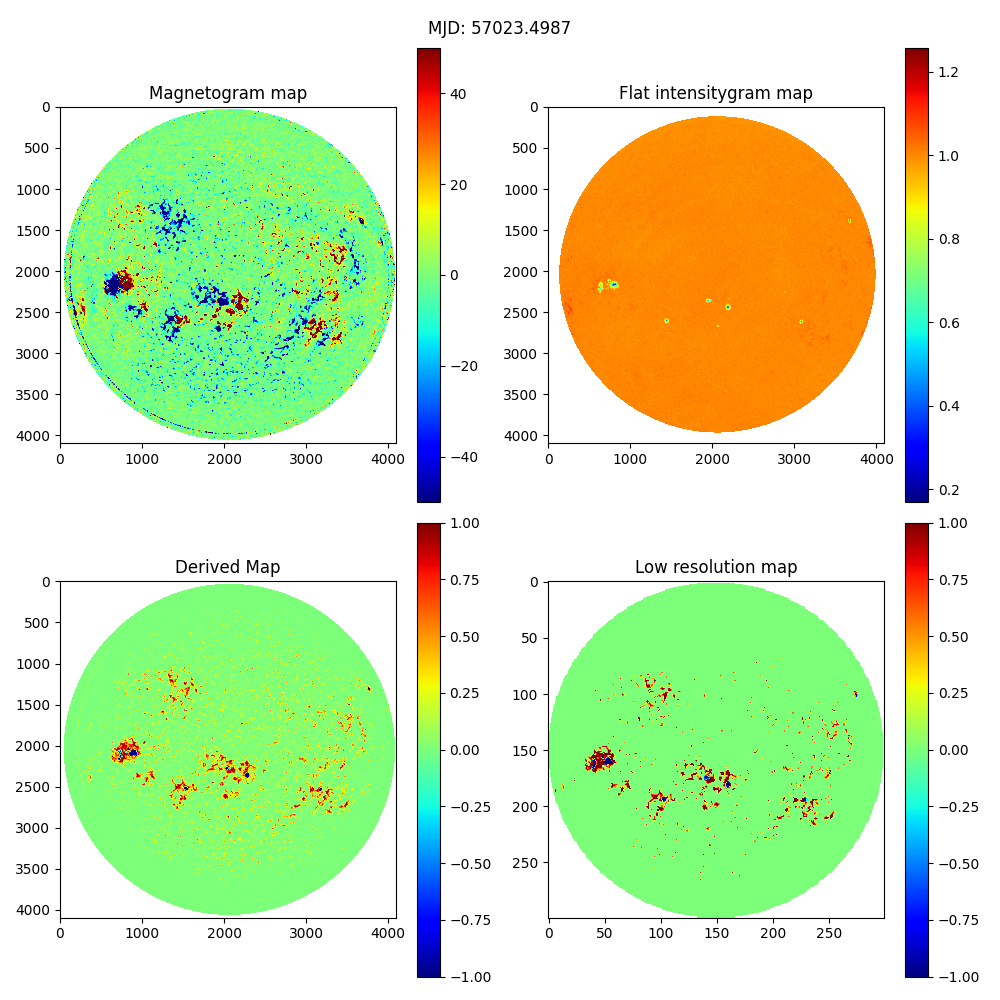}
   \caption{Pre-processing of SDO manetogram and flattened intensitygram images to derive active regions. Top left: Original SDO manetogram image. Regions with positive and negative magnetic fields are labeled in red and blue, respectively. Top right: Original SDO flattened intensitygram image. Regions with high contrast (in green) indicate the presence of spots. Bottom left: Derived high-resolution active region location with a dimension of 3900 $\times$ 3900. The facula regions are in red, while the spots are in blue. Bottom right: Low-resolution map of active regions with a size of 300 $\times$ 300, used as input of SOAP-GPU to speed up computation.}
   \label{fig_sod_img}%
   \end{figure*}

\cite{Haywood-2016MNRAS} extracted the solar active region properties using the dopplergram, intensity continuum, and magnetogram images from the Helioseismic and Magnetic Imager \citep[HMI,][]{Schou-2012SoPh}) on board SDO. We followed a similar approach to generate the active region map as input of SOAP-GPU. Since limb darkening is already included in SOAP-GPU, we only used the flatted intensitygram data to localize the spot regions and used the magnetogram map to localize the facula regions, as shown in the top panels of Fig.~\ref{fig_sod_img}. The original SDO FITS files were downloaded from the Joint Science Operations Center (JSOC).\footnote{\url{http://jsoc.stanford.edu/ajax/lookdata.html}} We used the magnetogram and flattened intensitygram data with a cadence of one day. We downloaded the data from January 1, 2015, to August 15, 2024, which covers the current HARPS-N solar observations (July 2015 to February 2024). The HARPS-N solar data that we used here are binned per day, and therefore we have one frame per day, for a total of 3494 active region maps.

Given that the original SDO data have a dimension of 3900 by 3900, it is computationally expensive for SOAP-GPU to use such high-resolution maps as input to perform the simulation. We therefore lowered the resolution of the flattened intensitygram and magnetogram from 3900 by 3900 to 300 by 300 for computational efficiency. We then detected spots as regions in the flattened intensitygram for which the contrast is less than 0.89 \citep{Haywood-2016MNRAS}. Regarding the detection of faculae, we adopted for the magnetogram the same threshold value of $24.0/\mu$ as mentioned in \cite{Haywood-2016MNRAS}. Pixels above this value were considered facula regions. The derived high-resolution active region map is illustrated in the bottom left panel of Fig.~\ref{fig_sod_img}. 

We lowered the resolution of the derived active region map from 3900 by 3900 to 300 by 300, which means that a subgrid of 13 by 13 in the high-resolution map is converted to 1 pixel in the low-resolution map. If more than 50\% of the low-resolution pixels were covered by a facula region in the high-resolution map, then the low-resolution pixel was considered to be covered by a facule. We repeated the same process to lower the resolution of the flat intensitygram. Once we derived both the low-resolution facula and spot maps at each timestamp, we combined the two into one active region map. If there were overlapping pixels, we counted them as spot pixels since spot regions are surrounded by facula regions. An example of a final low-resolution active region map used as input to SOAP-GPU is illustrated in the bottom right of Fig.~\ref{fig_sod_img}. 

We note that lowering the resolution of the SDO maps could have a impact on the final modeled RVs (see Sect.~\ref{sec3.2}) if active regions start to be truncated. However, we compared the derived RVs with the full and reduce resolution map of SDO and the results were extremely similar, demonstrating that the resolution that we chose was appropriate.

\subsection{Spectral time series results} \label{sec3.2}

\begin{figure*}[htbp]
  \centering
  \includegraphics[scale=0.3]{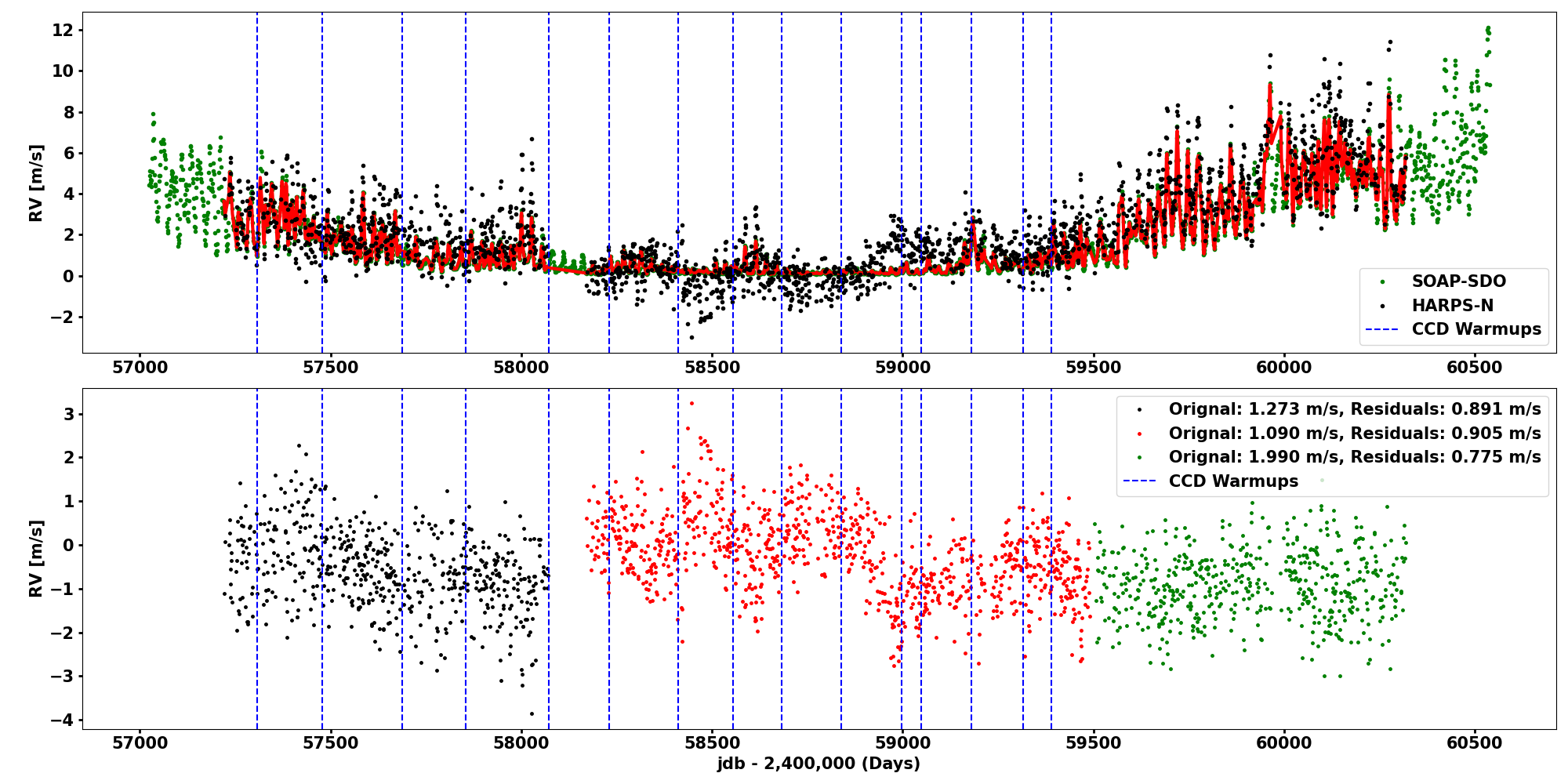}
  \caption{RV data of simulated spectral time series modeled with SDO images. The input seed spectra generated from the PHOENIX synthetic spectral library are used as the inputs of SOAP-GPU (see Sect.~\ref{sec2.2}). \emph{Top:} RV comparison between the SOAP-GPU simulated spectra and HARPS-N solar spectra. The RVs derived from the simulated spectra are highlighted in green while the RVs derived from HARPS-N solar spectra are shown in black. We interpolated the simulated RV time series (in red) to match the HARPS-N solar observation. The blue dashed vertical lines indicate the times of the CCD warm-ups. \emph{Bottom:} Residuals after subtracting simulated RVs from the HARPS-N solar RVs. The RV residuals are divided into three phases: the decreasing cycle 24 activity phase (black), the quiet phase (red), and the new increasing activity phase of cycle 25 (green). The corresponding rms are $0.891\,\rm{m/s}$, $0.905\,\rm{m/s}$, and $0.775\,\rm{m/s}$, respectively. The rms of HARPS-N solar spectra for these three phases are $1.273\,\rm{m/s}$, $1.090\,\rm{m/s}$, and $1.990\,\rm{m/s}$, and are shown here for comparison.}
    \label{Fig_rv_comparison}%
    \end{figure*}

We ran the SDO image-based simulation with the two sets of the input seed spectral cubes: the PHOENIX and the IAG cases. For each SDO active region map, there are three spectra simulated to account for the flux effect, inhibition of CB effect, and the total effect. The corresponding RV data points were obtained by using the ESPRESSO G2 mask. Compared with the simulation using only spot numbers, the simulation using the SDO maps contains accurate information about the active regions, which allows us to directly compare the results with the solar observation. We used the solar spectra observed by the HARPS-N solar telescope ranging from 2015 July 19 to 2024 January 8 \citep[][Dumusque et al. in prep.]{Dumusque-2021aa}. We collected 2040 daily binned solar spectra. All the solar spectra are preprocessed by the YARARA data pipeline \citep{Cretignier:2021aa} to remove systemic errors such as those induced by tellurics, ghosts, interference patterns, and stellar activity. Since we need to compare the simulated stellar activity with the observed one, we injected the YARARA stellar activity correction component back into the YARARA-cleaned spectra. We derived the RV time series of HARPS-N solar data by using the same ESPRESSO G2 mask as the simulated spectra. Since the SDO images and the HARPS-N solar observation have different sampling rates, we interpolate the simulated RV time series into the observation grid of HARPS-N solar data. The resulted RV time series for the PHEONIX and IAG cases are illustrated in Figs.~\ref{Fig_rv_comparison} and~\ref{Fig_rv_comparison_IAG}, respectively.

We divide the RV time series into three sections: the end of cycle 24 decreasing activity phase, the quiet phase, and the new cycle 25 increasing activity phase. By comparing the simulated RVs with the observed ones and deriving the RV rms by fitting the offset for each phase, we found for the decreasing and increasing activity phases that the RV rms can be strongly reduced from $1.27$ to $0.89$ and $1.99$ to $0.78\,\rm{m/s}$, respectively. For the quiet phase the RV rms is only marginally reduced from $1.09$ to $0.91\,\rm{m/s}$ which is not surprising as the SDO model is rather flat during this time due to the weak solar activity. This demonstrates that our SOAP-GPU SDO simulation is able to model well the observed RVs as stellar activity is mitigated to below 1 m/s level during the active phases. Regarding the increasing activity phase, we see that the residuals after correction of the SDO model are at the level of $0.78\,\rm{m/s}$, which is comparable to what is predicted from supergranulation \citep[e.g., $0.86$ and $0.68\,\rm{m/s}$ in][respectively]{Lakeland:2024aa, AlMoulla:2023aa}. For the decreasing activity phase, the RV rms residual is $0.89\,\rm{m/s}$, which is slightly higher than what is predicted from other stellar signals and the residuals clearly show a trend that is likely responsible for this extra jitter (removing this trend lowers the RV rms to $0.84\,\rm{m/s}$). We do not know the origin of this trend; however, it could be due to a different convective blueshift from one cycle to the next, or a different strength in its inhibition, which is a constant value equal to $315\,\rm{m/s}$ in our model. Finally, the RV rms of the residuals during the quiet phase is also greater than that predicted from supergranulation, suggesting that instrumental systematics are clearly at play. Two rather strong deviations compared to the mean RV around BJD 2458500 and 2459000 in the quiet phase are caused by a HARPS-N detector warm-up and by the change of the main ThAr lamp used for wavelength solution, respectively. The first intervention slightly changes the point spread function on the detector and induces a small RV shift; the second slightly changes the thorium spectrum, therefore affecting the wavelength solution and thus also inducing a RV offset. In addition, the gap in the data around BJD=2458000 corresponds to the fiber injecting light from the solar telescope to the calibration unit being damaged. After replacing it, we observed an RV offset of about 0.5 to 1 m/s. 
As the problem happened before injection into the calibration unit, it is difficult to understand how a different injection at this level could introduce an RV offset as in principle the octagonal fibers used by HARPS-N, in addition to a double scrambler, should strongly mitigate any difference in light-injection at the calibration unit level. However, a detector warm-up also happened within this observational gap and it could be the cause of this RV offset. When investigating the behavior of the RV around each warm-up, it seems that not all of them have a significant impact on the RVs. A detailed study on the effect of warm-ups is beyond the scope of this paper, but should be investigated in the future to understand precisely what is happening and hopefully find a way to mitigate the impact of warm-ups on RV precision.

\subsection{Comparison with HARPS-N data at the spectral level} \label{sec3.3}

The good agreement between the simulated spectra and the HARPS-N spectra in the RV space inspires us to further compare the two datasets at the spectral level. Given that the CCF is derived from a specific line selection, in our case from the ESPRESSO G2 mask, only those lines contribute to the corresponding RV value. Because the RV signal is measured from the spectral lines and not the stellar continuum, we decided to investigate the morphology of spectral lines in the two datasets rather than to compare the full spectra. Here we used the line list derived from \cite{Cretignier:2020aa}. Blended lines and weak lines were excluded from the analysis due to contamination and noise. We only used lines that are deeper than $0.1$ in normalized flux. For each selected spectral line, we compared in the model and in the real observations how the line Doppler shift and depth behave over time.

\begin{figure*}[htbp]
  \centering
  \includegraphics[scale=0.35]{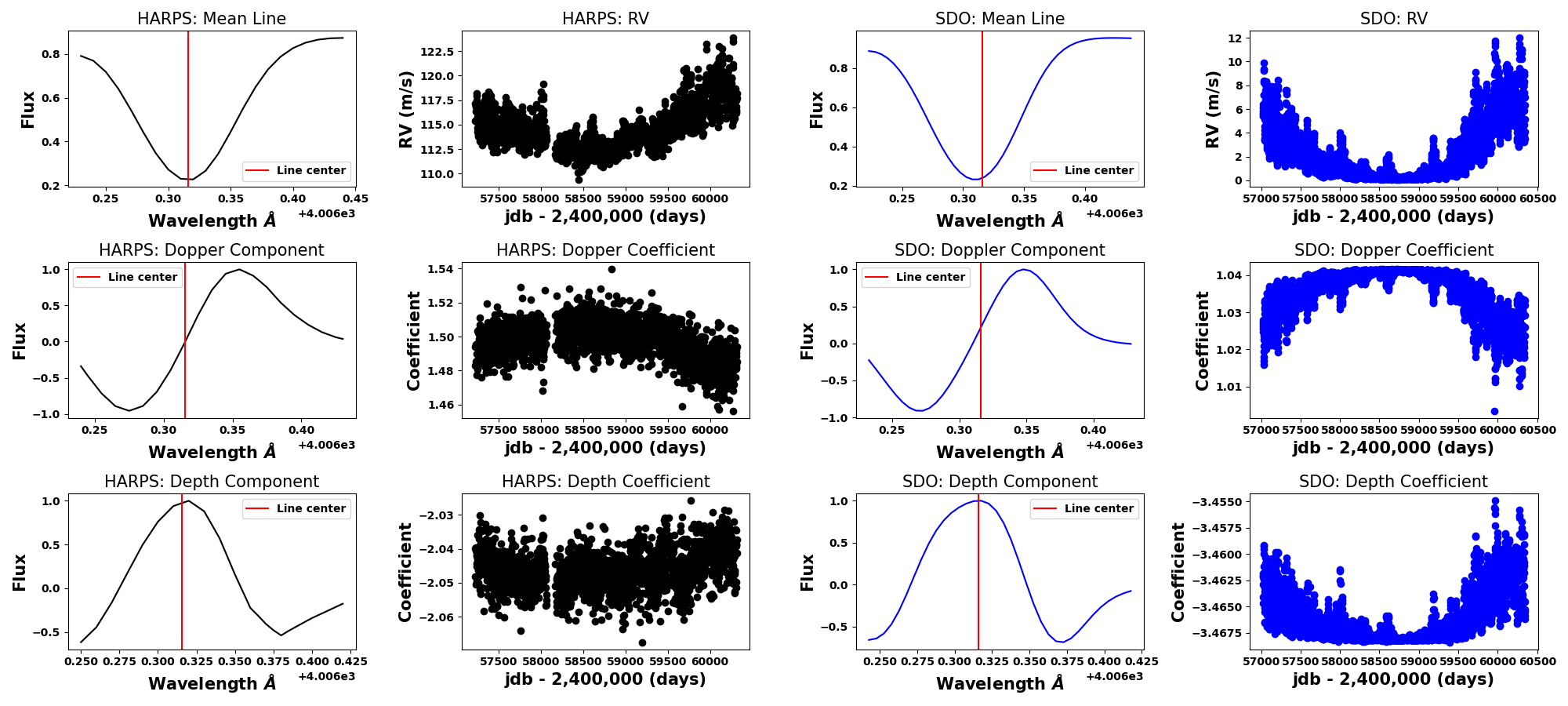}
  \caption{Example of line profile decomposition in the Doppler-shift and line-depth domains for the FeI line at 4006.31\,$\AA$. \emph{First column:} Line profiles of HARPS-N solar spectra. The top panel is the average line profile. The middle panel is the Doppler component of that line. The bottom panel is the depth component of the line. \emph{Second column:} Top panel is the RVs measured from the CCFs. The coefficients associated with the Doppler and depth components are shown in the middle and bottom panels, respectively. \emph{Third and fourth columns:} Same as the first and second columns but for the simulated spectral line. The Pearson correlation between the line in HARPS-N solar spectra and the same line in the simulated spectra is 0.70 for the Doppler component and 0.41 for the depth component}
    \label{Fig_line_analysis}%
    \end{figure*}  
    
\begin{figure*}[htbp]
  \centering
  \includegraphics[scale=0.35]{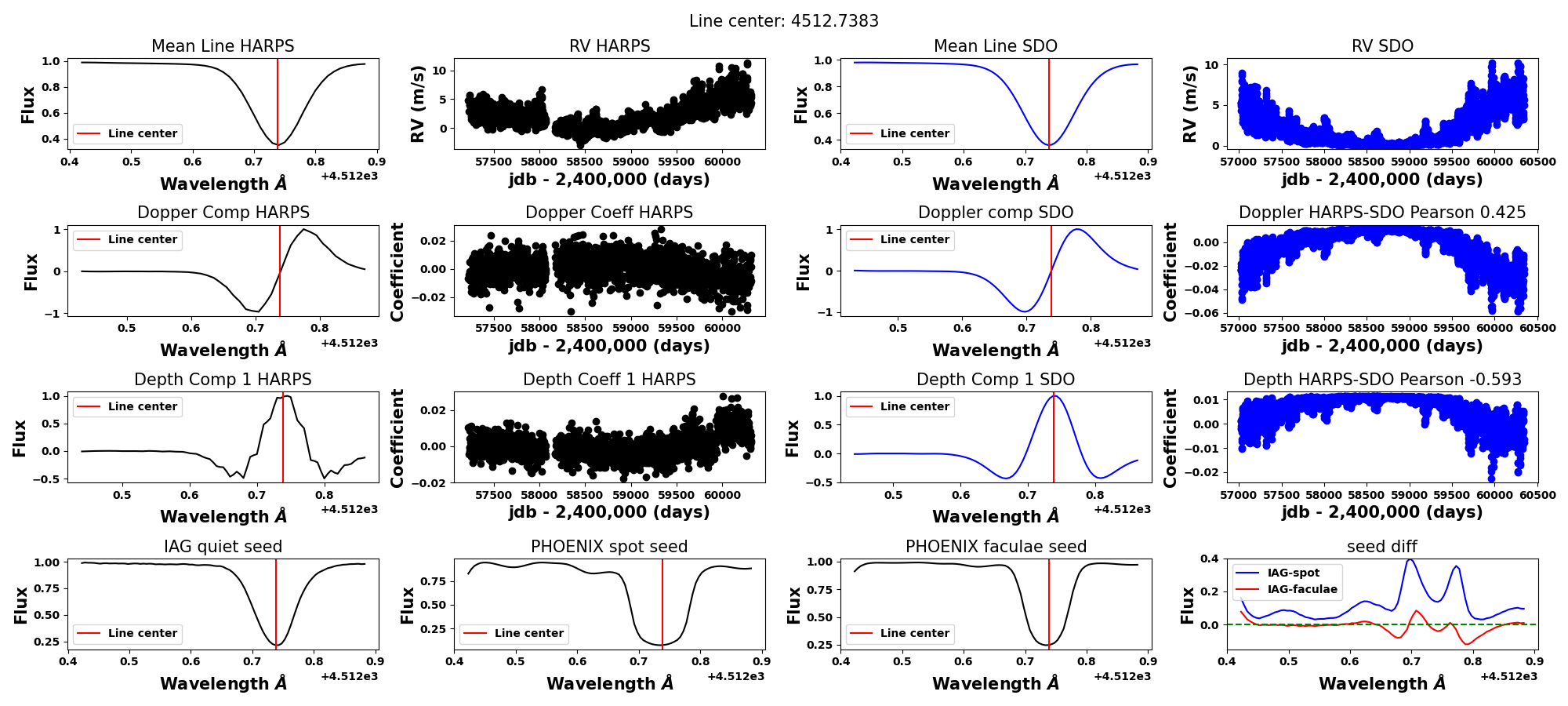}
  \caption{Similar to Fig.~\ref{Fig_line_analysis}, but for the FeI line at 4512.74\,$\AA$ from the IAG case. It has a negative Pearson correlation coefficient value of -0.593 for the depth coefficients. An additional row is added to highlight the input seed spectral line profile used in this simulation. \emph{Fourth row:} Line profile of the IAG spectrum at $\mu = 1.0$ for modeling the quiet Sun and line profiles from the PHOENIX spectral library for modeling spot and facula regions. The last panel in this row shows the line profile difference between the input seed spectrum of the quiet Sun and the input seed spectra of the active regions. A horizontal green dashed line indicates the zero level. The profile of $\Delta \mathbf{S}_{bconv, quiet-faculae}$ has regions greater than zero, which is not expected.}
    \label{Fig_line_decompostion_IAG}%
    \end{figure*}  

\begin{figure*}[htbp]
  \centering
  \includegraphics[scale=0.35]{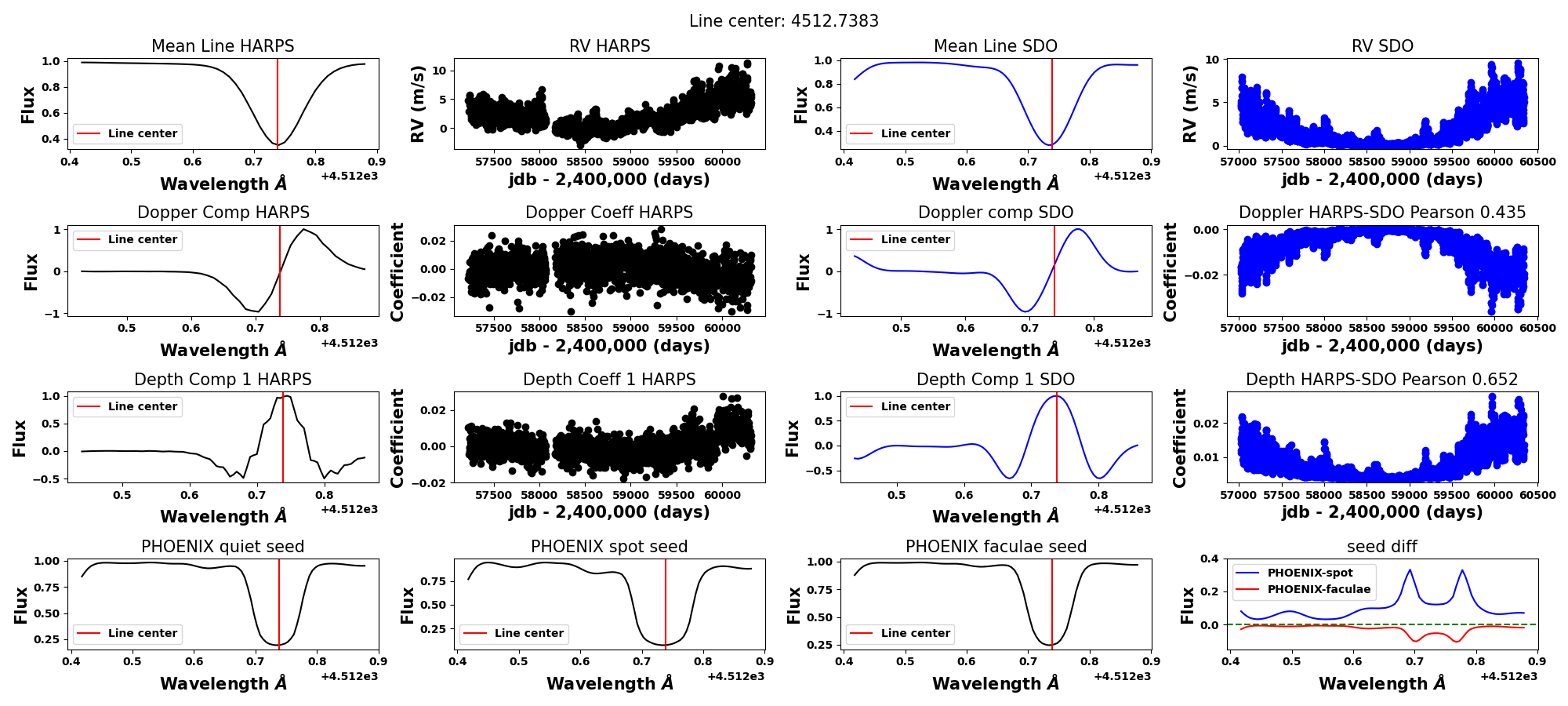}
  \caption{Same as Fig.~\ref{Fig_line_decompostion_IAG}, for the same line, but for the PHOENIX case. The correlation of depth coefficient shows a positive Pearson correlation coefficient value of 0.652 here. The last panel in the fourth row shows the difference between the input seed spectrum of quiet region. The profile of $\Delta \mathbf{S}_{bconv, quiet-faculae}$ is always below zero as expected.}
    \label{Fig_line_decompostion_PHOENIX}%
    \end{figure*}

    \begin{figure*}[htbp]
    \centering 
    \includegraphics[scale=0.35]{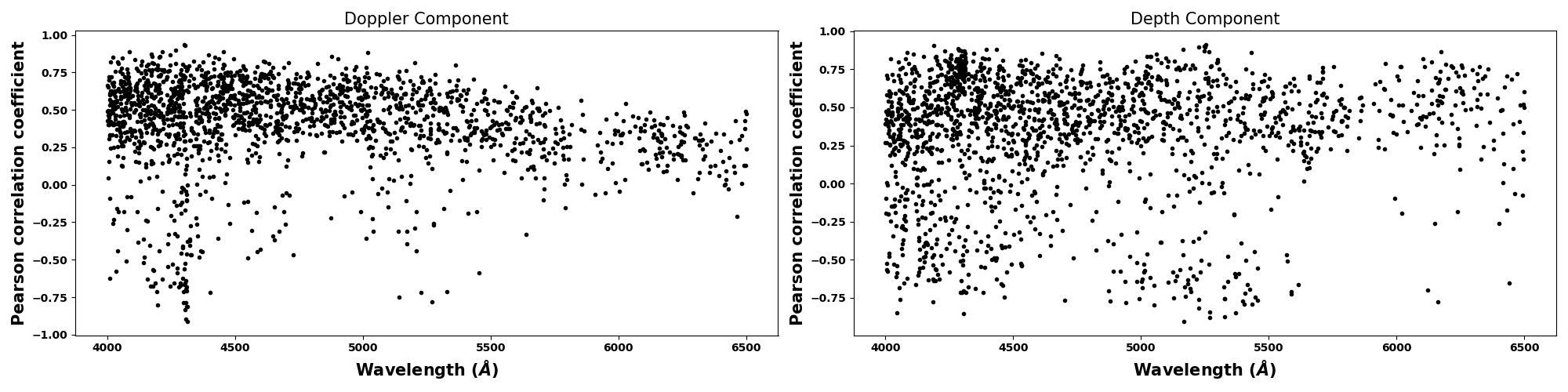}
    \caption{Line similarity in the wavelength domain for the SDO-based SOAP-GPU simulation using the PHOENIX input seed spectra (see Sect.~\ref{sec2.2}). \emph{Left:} Similarity of line profiles associated with the Doppler component. Line similarity decreases with wavelength, and therefore it seems that the SDO-based SOAP-GPU simulation can model blue lines better than red lines. This is likely due to telluric line contamination at the red part of the HARPS-N solar spectra. \emph{Right:} Similarity of line profiles associated with the depth component. There is no correlation between the Pearson correlation coefficient value and wavelength.}
    \label{Fig_LBL_wave_SDO}%
    \end{figure*}

    \begin{figure*}[htbp]
    \centering 
    \includegraphics[scale=0.35]{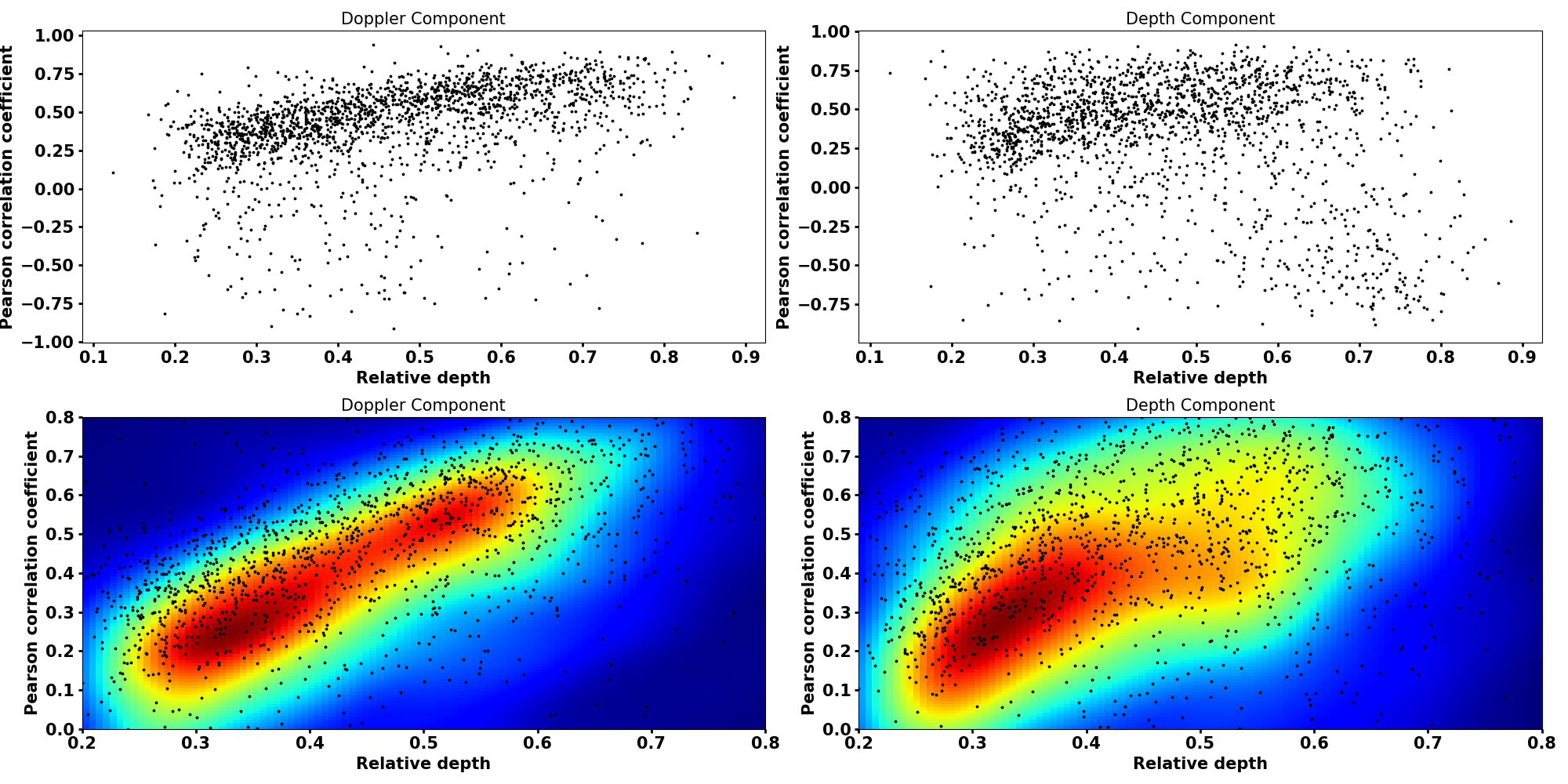}
    \caption{Line similarity in the relative line depth domain for the SDO-based SOAP-GPU simulation using the PHOENIX input seed spectra (see Sect.~\ref{sec2.2}).\emph{Top left:} Similarity of line profiles associated with the Doppler component. \emph{Bottom left:} Similar to the top left panel, but displaying only lines with Pearson correlation coefficient values greater than 0.0. Kernel density estimation is used for better visualization of the data distribution. There is a strong linear correlation between the line profile similarity and the relative line depth as the strong lines show higher similarity than the weak lines. This is likely because the strong lines have a higher signal-to-noise ratio than the weak lines. \emph{Top right:} Similarity of line profiles associated with the depth component. \emph{Bottom right:} Similar to the top right panel, but displaying only lines with Pearson correlation coefficient values greater than 0.0. Kernel density estimation is used for better visualization of the data distribution. A strong linear correlation is visible at relative line depth smaller than 0.4, where the lines are photon-limited. Beyond 0.4, the data is more uniformly distributed.}
    \label{Fig_LBL_coeff_SDO}%
    \end{figure*}

To investigate the line profile in the line Doppler shift and line shape domains, we first extracted the line shift and line depth basis that was then used to derive the coefficient of the line profile projected onto this basis. As the Doppler shift is proportional to the first derivative \citep[e.g.,][]{Bouchy-2001b}, we used the first-order derivative of the mean line profile as the Doppler shift basis. Regarding the line-depth basis, we can show that the second-order derivative is sensitive to line depth changes. Therefore, we used the second-order derivative of the mean line profile as the line-depth basis. In the first and the third columns of Fig.~\ref{Fig_line_analysis}, we show in the case of one spectral line the mean line profile as well as the line shift basis and line depth basis.

In order to quantify the line shift and line depth changes for a given spectral line as a function of time, we projected the line profile time series onto the line shift and line depth basis and looked at the obtained respective coefficients. We show the coefficient time series for the FeI line at 4006.31\,$\AA$ in the second and fourth columns of Fig.~\ref{Fig_line_analysis}. In both the HARPS-N and the SDO-based SOAP-GPU simulated spectral time series, the line shift coefficients and the line depth coefficients time series show a strong correlation with the corresponding RV time series. This is not surprising since stellar activity modifies the line shape, which can be decomposed as a shift and depth variation. To measure the similarity in the variation of line morphology between the HARPS-N spectra and the SDO-based SOAP-GPU spectra, we first computed for each spectral line and for the real and simulated cases the time series of the coefficients obtained when projecting onto the line shift or the depth basis. We then computed for each spectral line the Pearson correlation coefficient between the obtained coefficient time series. A line with a large positive Pearson correlation coefficient value indicates that its shift and/or depth variations are well modeled by SOAP-GPU. Some lines, however, present large negative Pearson correlation coefficient values, for example the line at 4512.74\,$\AA$ shown for both the IAG case and PHOENIX case in the third row of Fig.~\ref{Fig_line_decompostion_IAG} and Fig.~\ref{Fig_line_decompostion_PHOENIX}, respectively. The Pearson correlation coefficient value of the depth coefficient for the IAG case shows a strong anti-correlation of -0.593, while the correlation is 0.652 for the PHOENIX case. Given that the inhibition of convective blueshift is the main contribution to solar activity \citep[e.g.,][]{Meunier-2010a,Liebing:2021aa}, this anti-correlation is likely due to the spectral difference between the input seed spectrum of the quiet Sun and the input seed spectrum of active regions, as described by Equation 2 and Equation 10 in \cite{Zhao-2023A&A}. From Fig.~\ref{fig_sod_img} we know that facula regions are much larger than spot regions, making faculae the main contributor to the convective blueshift inhibition. The spectral difference between the input seed spectrum of the quiet Sun at $\mu = 1.0$ and the input seed spectrum of a facula (herefater $\Delta \mathbf{S}_{bconv, quiet-faculae}$) is illustrated in the last panels of the fourth row in Fig.~\ref{Fig_line_decompostion_IAG} and Fig.~\ref{Fig_line_decompostion_PHOENIX}. Faculae are hotter than the quiet Sun, and therefore we expect the spectral lines to be shallower due to the higher temperature. The value of $\Delta \mathbf{S}_{bconv, quiet-faculae}$ should therefore be smaller than zero. This is the case for the SOAP-GPU simulation using as input for the quiet Sun the PHOENIX synthetic library, as shown in Fig.~\ref{Fig_line_decompostion_PHOENIX}. However, for the IAG case in Fig.~\ref{Fig_line_decompostion_IAG}, $\Delta \mathbf{S}_{bconv, quiet-faculae}$ for the line at 4512.74\,$\AA$ goes above zero near the line core. Given that the temperature difference between the quiet region and the facula region is smaller than 250K \citep[250K at the limb, 33K at disk center][]{Meunier-2010a}, there is an inconsistency between the IAG observations and PHOENIX synthetic library for this small temperature difference. In other words, a mismatch between the line gradient and the effective temperature gradient for certain lines will lead to an anti-correlation between the real and simulated cases. It is therefore dangerous to mix input spectra seeds comings from real observations and synthetic libraries, and for that reason, as long as we do not have good observations for spots and faculae, we do not recommend using the IAG input seed spectra with SOAP-GPU.

We investigated the line similarity in different physical domains. We first investigated the line similarity in the wavelength domain, as shown in Fig.~\ref{Fig_LBL_wave_SDO} and Fig.~\ref{Fig_LBL_wave_SDO_IAG}. We find that there is a linear trend between the Pearson correlation coefficient value and the wavelength for the line shift component. As the wavelength increases, the Pearson correlation coefficient value decreases, as shown in the left panels of Fig.~\ref{Fig_LBL_wave_SDO} and Fig.~\ref{Fig_LBL_wave_SDO_IAG}. This means that the difference between the HARPS-N solar spectra and the SDO-based SOAP-GPU simulated spectra is larger in the red part of the spectrum than in the blue part. One possible explanation is the presence of telluric features. Since telluric lines start to pollute the observed HARPS-N spectra at a wavelength of $\sim$ 5000 \AA, the redder absorption lines suffer more from telluric line contamination than bluer spectral lines. In the line depth domain, the Pearson correlation coefficient values are uniformly distributed in the wavelength space and we cannot see any linear trend. One possible reason is that the contamination from telluric line affects the line shift more than the line depth as telluric lines contaminating any portion of a stellar spectral line will induce a Doppler shift, while only contamination at the precise location of the line core will affect line depth. As described in the previous paragraph, lines with negative Pearson correlation coefficient values indicate a mismatch between the line gradient and the effective temperature gradient.

We further investigated the line similarity in the line relative depth domain. The Pearson correlation coefficient value shows a linear trend with the line relative depth for the Doppler component. As the relative depth increases, the line similarity between the HARPS-N solar spectra and the SDO simulated spectra also increases for the line shift variation with only a small dispersion to negative Pearson correlation coefficient values, as shown in the left panels of Fig.~\ref{Fig_LBL_coeff_SDO}. For the IAG case, there is a strong correlation for lines with positive Pearson correlation coefficient values, as shown in the bottom left panel of Fig.~\ref{Fig_LBL_coeff_SDO_IAG}. Given the inconsistency between the IAG observation and PHOENIX synthetic library described above, the number of lines with negative Pearson correlation coefficient values in the IAG case is much larger than in the PHOENIX case. The possible explanation for this linear trend is that the morphology measurement on weak lines in the HARPS-N solar spectra is greatly affected by photon noise and the signal-to-noise ratio (S/N) of those weak lines is not sufficient to derive precise measurements. This is something that was also observed and explored in depth in the study of \citet{Cretignier:2020aa}.

For the line similarity of the depth component, there is also a strong correlation between the Pearson correlation coefficient value and the corresponding relative depth: only a small fraction of lines have negative Pearson correlation coefficient values, as shown in the top right panel of Fig.~\ref{Fig_LBL_coeff_SDO}. However, when lines with negative values are excluded, we found that the correlation is not linear across the entire relative depth range when the input seed spectra for SOAP-GPU are based on the PHOENIX library, as shown in the bottom right panel of Fig.~\ref{Fig_LBL_coeff_SDO}. For lines with a relative depth less than 0.4, there is a strong correlation between the relative depth and the line morphology, which is likely due, as in the case of the line shift, to the low S/N of weak lines in the HARPS-N solar spectra. For lines with relative depth greater than 0.4, in the PHOENIX case (see ~Fig.~\ref{Fig_LBL_coeff_SDO}), the slope of this linear trend becomes smaller and the data are more widely distributed. When the input seed spectra for SOAP-GPU are based on the IAG solar atlas, we found that both lines with positive and negative Pearson correlation coefficient values show a strong correlation (see the top right of Fig.~\ref{Fig_LBL_coeff_SDO_IAG}), similar to the case for the wavelength domain. After excluding lines with negative Pearson correlation coefficient values, we observed a strong correlation with line depth for all lines, again pointing, as in the case of the shift analysis, to a S/N dependence, as shown in the bottom right of Fig.~\ref{Fig_LBL_coeff_SDO_IAG}. The significant difference observed between the PHEONIX and the IAG simulated spectral time series for lines with positive Pearson correlation coefficient values, likely points toward a modeling of the depth of stellar lines that is not fully correct in the PHOENIX case. As explained in Sect.~\ref{sec2.2}, PHOENIX spectra do not include the proper convective blueshift as is naturally the case for the IAG solar atlas. Therefore, we artificially added convective blueshift by modifying the line bisector based on a few solar line measurements \citep{Lohner-Bottcher:2019aa}. We therefore assume that all lines have the same bisector, which is probably correct to first order for unblended lines, but will not work for blended lines. Although we make a careful selection of lines to reject evident blends, to a certain extent all lines are blended, and therefore having an impact on the way we model the line bisector.

\section{Discussion and conclusion}\label{sec4}

In this paper we demonstrated two methods of using SOAP-GPU to simulate realistic spectral time series affected by stellar activity perturbations for the Sun, and publish the simulations for the benefit of the community working on stellar activity mitigation in RV measurements. SOAP-GPU allows users to simulate solar spectral time series using as input the number of spots as the function of time. Other physical processes, for example the active size evolution curve, the spatial migration, and the initial size distribution, are parameterized by empirical equations derived from previous solar observations. The input spectral cube used to model the quiet solar surface as a function of position on the limb ($\mu$ angle) can either come from PHOENIX synthetic spectra on which the bisector of spectral line is modified as a function of $\mu$, based on solar observations from \citet{Lohner-Bottcher:2019aa}, or from observations of the full visible spectrum taken at different heliocentric positions \citep[][]{Ellwarth-2023aa}.

By using the spot number as a function of time to model the spectral time series affected by solar activity, the prominent features of solar activity can be well modeled. The long-term magnetic cycle and the signals associated with solar rotation can clearly be observed in the derived RVs from the simulated spectra (see Fig.~\ref{Fig_RV_periodogram}). We are therefore confident that such a strategy can be used to realistically model stellar activity on stars other than the Sun. However, when going far from the Sun in spectral type, although the use of PHOENIX spectra and the modification of the convective blueshift can adapt to different spectral types \citep[see][]{Zhao-2023A&A}, we should be careful that the active region properties (initial size, decay rate, latitude migration, and the effective temperatures of active regions) are based on solar observations and may differ for other spectral types. 

In order to precisely model solar activity and compare with the corresponding HARPS-N solar spectra, we can also give as input to SOAP-GPU the magnetograms and flat intensity maps from SDO observations. Unlike generating the location of active regions with the spot number curve and empirical equations from solar observations, the active region location and size at different timestamps can be precisely extracted, which allows us to compare on a one-to-one basis the simulated spectra with the HARPS-N solar observations. 

In the RV space, the time series of SDO simulated spectra is in good agreement with the RVs derived from the HARPS-N solar spectra, with a residual RV rms of $\sim 0.8-0.9\,\rm{m/s}$. In our analysis, we split the time series into three parts, two corresponding to active phases, and one to the inactive phase. We clearly demonstrate that our SOAP-GPU SDO model is able to significantly mitigate stellar activity during the decreasing and increasing active phases to levels below the level of 1 m/s, from 1.27 and $1.99\,\rm{m/s}$ down to 0.89 and $0.78\,\rm{m/s}$, respectively. While the increasing active phase show a RV rms of 0.78 m/s similar to what is expected from supergranulation, the decreasing phase still show higher jitter, 0.89 m/s, likely due to a trend in the residuals, of unknown origin. We also see some significant residuals during the inactive phase (RV rms = 0.91 m/s), which is likely due to residual instrumental systematics. A detailed study of those systematics is beyond the scope of this paper, but the initial investigation points toward warm-ups of the detector and change of the master thorium-argon lamp used for the wavelength solution. 

At the spectral level, we compared spectral lines in the line shift and line depth domain. We measured the similarity in line morphology variation between the two spectral datasets by projecting each line onto a shift and depth basis and comparing the Pearson correlation between the obtained coefficients. Some of the lines show negative Pearson correlation coefficient values, indicating a mismatch between the line gradient and the effective temperature gradient. Although the necessary physics are naturally included in the IAG observations, the number of lines with negative Pearson correlation coefficient values in the IAG case is larger than in the PHOENIX case. This is due to an inconsistency between the IAG observations and the PHOENIX synthetic library which is always used to model the spectra of active regions. For this reason, we do not recommend that users adopt the IAG spectra as input of SOAP-GPU as long as we do not have good observations of spots and faculae to use for modeling active regions. It is therefore crucial in the future to obtain active region observations at different heliocentric positions covering the full visible spectral range. Future solar observations using the ESPRESSO spectrograph may help us in that direction \citep{Santos-2023spfi}.

After we filtered out lines with negative Pearson correlation coefficient values, we found that the line similarity of the Doppler component decreases with wavelength, which is likely due to the contamination of the telluric lines in the HARPS-N solar spectra. The line similarity of the Doppler component also increases with the line relative depth. This strong correlation is likely due to the low S/N of weak lines in the HARPS-N solar spectra that decreases the observed correlation \citep[e.g.,][]{Cretignier:2020aa}. In the case of the line depth component, the line similarity also shows some correlation with relative line depth and the same explanation related to low S/N can be given as well.

We are conscious that the use of a single spectral line bisector to represent all of them is a current limitation to this first SOAP-GPU solar activity simulation at the spectral level, however, we demonstrated that even with this simplification, the general features of the HARPS-N solar spectra, both at the RV level, but also at the spectral line-shape variation level, can be well modeled by SOAP-GPU when using the SDO images as input. We are therefore confident that SOAP-GPU is able to satisfactorily model stellar activity on the Sun and therefore the published spectral time series should be considered as a useful test dataset for evaluating the performance of any stellar activity mitigating technique working at the spectral level. This does not guarantee that the same techniques working well on this dataset will work equally well on real observations as instrumental signals, but also other stellar signals such as supergranulation are not included in SOAP-GPU. This first SOAP-GPU simulation of solar activity at the spectral level could easily be improved further in terms of a realism once either IAG spectra for spots and faculae at different $\mu$ angles, or MuRAM simulations modeling the full visible spectra of quiet and active regions at different $\mu$ angles are available. It will then be easy to use those as input of SOAP-GPU to produce a more realistic solar simulation.

\begin{acknowledgements}
We thank the anonymous referee for the insightful and constructive feedback on this paper. X.D acknowledges the support from the European Research Council (ERC) under the European Union’s Horizon 2020 research and innovation programme (grant agreement SCORE No 851555) and from the Swiss National Science Foundation under the grant SPECTRE (No 200021\_215200). This work has been carried out within the framework of the NCCR PlanetS supported by the Swiss National Science Foundation under grants 51NF40\_182901 and 51NF40\_205606.

\end{acknowledgements}

\bibliographystyle{aa}
\bibliography{Yinan_Zhao_bibli}

\begin{thebibliography}{44}
\expandafter\ifx\csname natexlab\endcsname\relax\def\natexlab#1{#1}\fi

\bibitem[{{Al Moulla} {et~al.}(2024){Al Moulla}, {Dumusque}, \&
  {Cretignier}}]{AlMoulla:2024aa}
{Al Moulla}, K., {Dumusque}, X., \& {Cretignier}, M. 2024, \aap, 683, A106

\bibitem[{{Al Moulla} {et~al.}(2022){Al Moulla}, {Dumusque}, {Cretignier},
  {Zhao}, \& {Valenti}}]{AlMoulla:2022aa}
{Al Moulla}, K., {Dumusque}, X., {Cretignier}, M., {Zhao}, Y., \& {Valenti},
  J.~A. 2022, \aap, 664, A34

\bibitem[{{Al Moulla} {et~al.}(2023){Al Moulla}, {Dumusque}, {Figueira}, {Lo
  Curto}, {Santos}, \& {Wildi}}]{AlMoulla:2023aa}
{Al Moulla}, K., {Dumusque}, X., {Figueira}, P., {et~al.} 2023, \aap, 669, A39

\bibitem[{{Baumann} \& {Solanki}(2005)}]{Baumann-2005aa}
{Baumann}, I. \& {Solanki}, S.~K. 2005, \aap, 443, 1061

\bibitem[{{Berdyugina} \& {Usoskin}(2003)}]{Berdyugina:2003aa}
{Berdyugina}, S.~V. \& {Usoskin}, I.~G. 2003, \aap, 405, 1121

\bibitem[{{Borgniet} {et~al.}(2015){Borgniet}, {Meunier}, \&
  {Lagrange}}]{Borgniet-2015aa}
{Borgniet}, S., {Meunier}, N., \& {Lagrange}, A.~M. 2015, \aap, 581, A133

\bibitem[{{Bouchy} {et~al.}(2001){Bouchy}, {Pepe}, \& {Queloz}}]{Bouchy-2001b}
{Bouchy}, F., {Pepe}, F., \& {Queloz}, D. 2001, \aap, 374, 733

\bibitem[{{Cavallini} {et~al.}(1985){Cavallini}, {Ceppatelli}, \&
  {Righini}}]{Cavallini-1985a}
{Cavallini}, F., {Ceppatelli}, G., \& {Righini}, A. 1985, \aap, 143, 116

\bibitem[{{Collier Cameron} {et~al.}(2021){Collier Cameron}, {Ford}, {Shahaf},
  {Aigrain}, {Dumusque}, {Haywood}, {Mortier}, {Phillips}, {Buchhave},
  {Cecconi}, {Cegla}, {Cosentino}, {Cr{\'e}tignier}, {Ghedina}, {Gonz{\'a}lez},
  {Latham}, {Lodi}, {L{\'o}pez-Morales}, {Micela}, {Molinari}, {Pepe},
  {Piotto}, {Poretti}, {Queloz}, {Juan}, {S{\'e}gransan}, {Sozzetti},
  {Szentgyorgyi}, {Thompson}, {Udry}, \& {Watson}}]{Collier-Cameron:2021aa}
{Collier Cameron}, A., {Ford}, E.~B., {Shahaf}, S., {et~al.} 2021, \mnras, 505,
  1699

\bibitem[{{Collier Cameron} {et~al.}(2019){Collier Cameron}, {Mortier},
  {Phillips}, {Dumusque}, {Haywood}, {Langellier}, {Watson}, {Cegla}, {Costes},
  {Charbonneau}, {Coffinet}, {Latham}, {Lopez-Morales}, {Malavolta},
  {Maldonado}, {Micela}, {Milbourne}, {Molinari}, {Saar}, {Thompson},
  {Buchschacher}, {Cecconi}, {Cosentino}, {Ghedina}, {Glenday}, {Gonzalez},
  {Li}, {Lodi}, {Lovis}, {Pepe}, {Poretti}, {Rice}, {Sasselov}, {Sozzetti},
  {Szentgyorgyi}, {Udry}, \& {Walsworth}}]{Collier-Cameron-2019MNRAS}
{Collier Cameron}, A., {Mortier}, A., {Phillips}, D., {et~al.} 2019, \mnras,
  487, 1082

\bibitem[{{Crass} {et~al.}(2021){Crass}, {Gaudi}, {Leifer}, {Beichman},
  {Bender}, {Blackwood}, {Burt}, {Callas}, {Cegla}, {Diddams}, {Dumusque},
  {Eastman}, {Ford}, {Fulton}, {Gibson}, {Halverson}, {Haywood}, {Hearty},
  {Howard}, {Latham}, {L{\"o}hner-B{\"o}ttcher}, {Mamajek}, {Mortier},
  {Newman}, {Plavchan}, {Quirrenbach}, {Reiners}, {Robertson}, {Roy}, {Schwab},
  {Seifahrt}, {Szentgyorgyi}, {Terrien}, {Teske}, {Thompson}, \&
  {Vasisht}}]{Crass:2021aa}
{Crass}, J., {Gaudi}, B.~S., {Leifer}, S., {et~al.} 2021, arXiv e-prints,
  arXiv:2107.14291

\bibitem[{{Cretignier} {et~al.}(2020){Cretignier}, {Dumusque}, {Allart},
  {Pepe}, \& {Lovis}}]{Cretignier:2020aa}
{Cretignier}, M., {Dumusque}, X., {Allart}, R., {Pepe}, F., \& {Lovis}, C.
  2020, \aap, 633, A76

\bibitem[{{Cretignier} {et~al.}(2021){Cretignier}, {Dumusque}, {Hara}, \&
  {Pepe}}]{Cretignier:2021aa}
{Cretignier}, M., {Dumusque}, X., {Hara}, N.~C., \& {Pepe}, F. 2021, \aap, 653,
  A43

\bibitem[{{Cretignier} {et~al.}(2022){Cretignier}, {Dumusque}, \&
  {Pepe}}]{Cretignier-2022aa}
{Cretignier}, M., {Dumusque}, X., \& {Pepe}, F. 2022, \aap, 659, A68

\bibitem[{{de Beurs} {et~al.}(2022){de Beurs}, {Vanderburg}, {Shallue},
  {Dumusque}, {Cameron}, {Leet}, {Buchhave}, {Cosentino}, {Ghedina}, {Haywood},
  {Langellier}, {Latham}, {L{\'o}pez-Morales}, {Mayor}, {Micela}, {Milbourne},
  {Mortier}, {Molinari}, {Pepe}, {Phillips}, {Pinamonti}, {Piotto}, {Rice},
  {Sasselov}, {Sozzetti}, {Udry}, \& {Watson}}]{Beurs:2022aa}
{de Beurs}, Z.~L., {Vanderburg}, A., {Shallue}, C.~J., {et~al.} 2022, \aj, 164,
  49

\bibitem[{{Dumusque}(2018)}]{Dumusque-2018aa}
{Dumusque}, X. 2018, \aap, 620, A47

\bibitem[{{Dumusque} {et~al.}(2014){Dumusque}, {Boisse}, \&
  {Santos}}]{Dumusque-2014b}
{Dumusque}, X., {Boisse}, I., \& {Santos}, N.~C. 2014, \apj, 796, 132

\bibitem[{{Dumusque} {et~al.}(2021){Dumusque}, {Cretignier}, {Sosnowska},
  {Buchschacher}, {Lovis}, {Phillips}, {Pepe}, {Alesina}, {Buchhave},
  {Burnier}, {Cecconi}, {Cegla}, {Cloutier}, {Collier Cameron}, {Cosentino},
  {Ghedina}, {Gonz{\'a}lez}, {Haywood}, {Latham}, {Lodi}, {L{\'o}pez-Morales},
  {Maldonado}, {Malavolta}, {Micela}, {Molinari}, {Mortier}, {P{\'e}rez
  Ventura}, {Pinamonti}, {Poretti}, {Rice}, {Riverol}, {Riverol}, {San Juan},
  {S{\'e}gransan}, {Sozzetti}, {Thompson}, {Udry}, \&
  {Wilson}}]{Dumusque-2021aa}
{Dumusque}, X., {Cretignier}, M., {Sosnowska}, D., {et~al.} 2021, \aap, 648,
  A103

\bibitem[{{Dumusque} {et~al.}(2015){Dumusque}, {Glenday}, {Phillips},
  {Buchschacher}, {Collier Cameron}, {Cecconi}, {Charbonneau}, {Cosentino},
  {Ghedina}, {Latham}, {Li}, {Lodi}, {Lovis}, {Molinari}, {Pepe}, {Udry},
  {Sasselov}, {Szentgyorgyi}, \& {Walsworth}}]{Dumusque-2015ApJ}
{Dumusque}, X., {Glenday}, A., {Phillips}, D.~F., {et~al.} 2015, \apjl, 814,
  L21

\bibitem[{{Ellwarth} {et~al.}(2023){Ellwarth}, {Sch{\"a}fer}, {Reiners}, \&
  {Zechmeister}}]{Ellwarth-2023aa}
{Ellwarth}, M., {Sch{\"a}fer}, S., {Reiners}, A., \& {Zechmeister}, M. 2023,
  \aap, 673, A19

\bibitem[{{Feng} {et~al.}(2017){Feng}, {Tuomi}, {Jones}, {Barnes},
  {Anglada-Escud{\'e}}, {Vogt}, \& {Butler}}]{Feng:2017ac}
{Feng}, F., {Tuomi}, M., {Jones}, H.~R.~A., {et~al.} 2017, \aj, 154, 135

\bibitem[{{Gilbertson} {et~al.}(2020){Gilbertson}, {Ford}, \&
  {Dumusque}}]{Gilbertson:2020aa}
{Gilbertson}, C., {Ford}, E.~B., \& {Dumusque}, X. 2020, Research Notes of the
  American Astronomical Society, 4, 59

\bibitem[{{Gray}(2009)}]{Gray-2009}
{Gray}, D.~F. 2009, \apj, 697, 1032

\bibitem[{{Hathaway}(2011)}]{Hathaway-2011SoPh}
{Hathaway}, D.~H. 2011, \solphys, 273, 221

\bibitem[{{Haywood} {et~al.}(2016){Haywood}, {Collier Cameron}, {Unruh},
  {Lovis}, {Lanza}, {Llama}, {Deleuil}, {Fares}, {Gillon}, {Moutou}, {Pepe},
  {Pollacco}, {Queloz}, \& {S{\'e}gransan}}]{Haywood-2016MNRAS}
{Haywood}, R.~D., {Collier Cameron}, A., {Unruh}, Y.~C., {et~al.} 2016, \mnras,
  457, 3637

\bibitem[{{Husser} {et~al.}(2013){Husser}, {Wende-von Berg}, {Dreizler},
  {Homeier}, {Reiners}, {Barman}, \& {Hauschildt}}]{Husser-2013aa}
{Husser}, T.~O., {Wende-von Berg}, S., {Dreizler}, S., {et~al.} 2013, \aap,
  553, A6

\bibitem[{{Lakeland} {et~al.}(2024){Lakeland}, {Naylor}, {Haywood}, {Meunier},
  {Rescigno}, {Dalal}, {Mortier}, {Thompson}, {Cameron}, {Dumusque},
  {L{\'o}pez-Morales}, {Pepe}, {Rice}, {Sozzetti}, {Udry}, {Ford}, {Ghedina},
  \& {Lodi}}]{Lakeland:2024aa}
{Lakeland}, B.~S., {Naylor}, T., {Haywood}, R.~D., {et~al.} 2024, \mnras, 527,
  7681

\bibitem[{{Langellier} {et~al.}(2021){Langellier}, {Milbourne}, {Phillips},
  {Haywood}, {Saar}, {Mortier}, {Malavolta}, {Thompson}, {Collier Cameron},
  {Dumusque}, {Cegla}, {Latham}, {Maldonado}, {Watson}, {Buchschacher},
  {Cecconi}, {Charbonneau}, {Cosentino}, {Ghedina}, {Gonzalez}, {Li}, {Lodi},
  {L{\'o}pez-Morales}, {Micela}, {Molinari}, {Pepe}, {Poretti}, {Rice},
  {Sasselov}, {Sozzetti}, {Udry}, \& {Walsworth}}]{Langellier:2021aj}
{Langellier}, N., {Milbourne}, T.~W., {Phillips}, D.~F., {et~al.} 2021, \aj,
  161, 287

\bibitem[{{Liebing} {et~al.}(2021){Liebing}, {Jeffers}, {Reiners}, \&
  {Zechmeister}}]{Liebing:2021aa}
{Liebing}, F., {Jeffers}, S.~V., {Reiners}, A., \& {Zechmeister}, M. 2021,
  \aap, 654, A168

\bibitem[{{Lin} {et~al.}(2022){Lin}, {Monson}, {Mahadevan}, {Ninan},
  {Halverson}, {Nitroy}, {Bender}, {Logsdon}, {Kanodia}, {Terrien}, {Roy},
  {Luhn}, {Gupta}, {Ford}, {Hearty}, {Laher}, {Hunting}, {McBride}, {Salazar
  Rivera}, {Rajagopal}, {Wolf}, {Robertson}, {Wright}, {Blake}, {Ca{\~n}as},
  {Lubar}, {McElwain}, {Ramsey}, {Schwab}, \& {Stefansson}}]{Lin:2022aj}
{Lin}, A. S.~J., {Monson}, A., {Mahadevan}, S., {et~al.} 2022, \aj, 163, 184

\bibitem[{{L{\"o}hner-B{\"o}ttcher} {et~al.}(2019){L{\"o}hner-B{\"o}ttcher},
  {Schmidt}, {Schlichenmaier}, {Steinmetz}, \&
  {Holzwarth}}]{Lohner-Bottcher:2019aa}
{L{\"o}hner-B{\"o}ttcher}, J., {Schmidt}, W., {Schlichenmaier}, R.,
  {Steinmetz}, T., \& {Holzwarth}, R. 2019, \aap, 624, A57

\bibitem[{{Martinez Pillet} {et~al.}(1993){Martinez Pillet}, {Moreno-Insertis},
  \& {Vazquez}}]{Martinez-1993aa}
{Martinez Pillet}, V., {Moreno-Insertis}, F., \& {Vazquez}, M. 1993, \aap, 274,
  521

\bibitem[{{Mayor} \& {Queloz}(1995)}]{Mayor-1995Nature}
{Mayor}, M. \& {Queloz}, D. 1995, \nat, 378, 355

\bibitem[{{Meunier} {et~al.}(2010){Meunier}, {Desort}, \&
  {Lagrange}}]{Meunier-2010a}
{Meunier}, N., {Desort}, M., \& {Lagrange}, A.-M. 2010, \aap, 512, A39

\bibitem[{{Palumbo} {et~al.}(2024){Palumbo}, {Ford}, {Gonzalez}, {Wright}, {Al
  Moulla}, \& {Schlichenmaier}}]{Palumbo:2024aa}
{Palumbo}, M.~L., {Ford}, E.~B., {Gonzalez}, E.~B., {et~al.} 2024, \aj, 168, 46

\bibitem[{{Phillips} {et~al.}(2016){Phillips}, {Glenday}, {Dumusque},
  {Buchschacher}, {Collier Cameron}, {Cecconi}, {Charbonneau}, {Cosentino},
  {Ghedina}, {Haywood}, {Latham}, {Li}, {Lodi}, {Lovis}, {Molinari}, {Pepe},
  {Sasselov}, {Szentgyorgyi}, {Udry}, \& {Walsworth}}]{Phillips:2016aa}
{Phillips}, D.~F., {Glenday}, A.~G., {Dumusque}, X., {et~al.} 2016, in Society
  of Photo-Optical Instrumentation Engineers (SPIE) Conference Series, Vol.
  9912, Advances in Optical and Mechanical Technologies for Telescopes and
  Instrumentation II, 99126Z

\bibitem[{{Rubenzahl} {et~al.}(2023){Rubenzahl}, {Halverson}, {Walawender},
  {Hill}, {Howard}, {Brown}, {Ida}, {Tehero}, {Fulton}, {Gibson}, {Kassis},
  {Smith}, {Wold}, \& {Payne}}]{Rubenzahl-2023pasp}
{Rubenzahl}, R.~A., {Halverson}, S., {Walawender}, J., {et~al.} 2023, \pasp,
  135, 125002

\bibitem[{{Santos}(2023)}]{Santos-2023spfi}
{Santos}, N.~C. 2023, in Spectral Fidelity, 12

\bibitem[{{Schou} {et~al.}(2012){Schou}, {Scherrer}, {Bush}, {Wachter},
  {Couvidat}, {Rabello-Soares}, {Bogart}, {Hoeksema}, {Liu}, {Duvall}, {Akin},
  {Allard}, {Miles}, {Rairden}, {Shine}, {Tarbell}, {Title}, {Wolfson},
  {Elmore}, {Norton}, \& {Tomczyk}}]{Schou-2012SoPh}
{Schou}, J., {Scherrer}, P.~H., {Bush}, R.~I., {et~al.} 2012, \solphys, 275,
  229

\bibitem[{{SILSO World Data Center}(2024)}]{sidc}
{SILSO World Data Center}. 2024, International Sunspot Number Monthly Bulletin
  and online catalogue

\bibitem[{{Zhao} {et~al.}(2023){Zhao}, {Dumusque}, {Ford}, {Llama}, {Mortier},
  {Bedell}, {Al Moulla}, {Bender}, {Blake}, {Brewer}, {Collier Cameron},
  {Cosentino}, {Figueira}, {Fischer}, {Ghedina}, {Gonzalez}, {Halverson},
  {Kanodia}, {Latham}, {Lin}, {Lo Curto}, {Lodi}, {Logsdon}, {Lovis},
  {Mahadevan}, {Monson}, {Ninan}, {Pepe}, {Roettenbacher}, {Roy}, {Santos},
  {Schwab}, {Stef{\'a}nsson}, {Szymkowiak}, {Terrien}, {Udry}, {Weiss},
  {Wildi}, {Wildi}, \& {Wright}}]{Zhao:2023ab}
{Zhao}, L.~L., {Dumusque}, X., {Ford}, E.~B., {et~al.} 2023, \aj, 166, 173

\bibitem[{{Zhao} {et~al.}(2022){Zhao}, {Fischer}, {Ford}, {Wise}, {Cretignier},
  {Aigrain}, {Barragan}, {Bedell}, {Buchhave}, {Camacho}, {Cegla},
  {Cisewski-Kehe}, {Collier Cameron}, {de Beurs}, {Dodson-Robinson},
  {Dumusque}, {Faria}, {Gilbertson}, {Haley}, {Harrell}, {Hogg}, {Holzer},
  {John}, {Klein}, {Lafarga}, {Lienhard}, {Maguire-Rajpaul}, {Mortier},
  {Nicholson}, {Palumbo}, {Ramirez Delgado}, {Shallue}, {Vanderburg}, {Viana},
  {Zhao}, {Zicher}, {Cabot}, {Henry}, {Roettenbacher}, {Brewer}, {Llama},
  {Petersburg}, \& {Szymkowiak}}]{zhao:2022AJ}
{Zhao}, L.~L., {Fischer}, D.~A., {Ford}, E.~B., {et~al.} 2022, \aj, 163, 171

\bibitem[{{Zhao} \& {Dumusque}(2023)}]{Zhao-2023A&A}
{Zhao}, Y. \& {Dumusque}, X. 2023, \aap, 671, A11

\bibitem[{{Zhao} {et~al.}(2024){Zhao}, {Dumusque}, {Cretignier}, {Cameron},
  {Latham}, {L{\'o}pez-Morales}, {Mayor}, {Sozzetti}, {Cosentino},
  {G{\'o}mez-Vargas}, {Pepe}, \& {Udry}}]{zhao:2024aa}
{Zhao}, Y., {Dumusque}, X., {Cretignier}, M., {et~al.} 2024, \aap, 687, A281

\end{thebibliography}

\begin{appendix}
\onecolumn
\section{Solar spectra modeling using IAG spectra}\label{sec_App}

We summarize the results of solar activity simulations using IAG input seed spectra for the quiet solar disk. For modeling of active regions, we continue to use spectra from PHOENIX synthetic library with $\rm{T_{eff}} = 6028K$ for facula regions and $\rm{T_{eff}} = 5115K$ for spot regions. For both simulations, using either spot number or SDO data, we employ the same active regions map time series as in the PHOENIX case.

  \begin{figure}[htbp]
  \centering
  \includegraphics[scale=0.28]{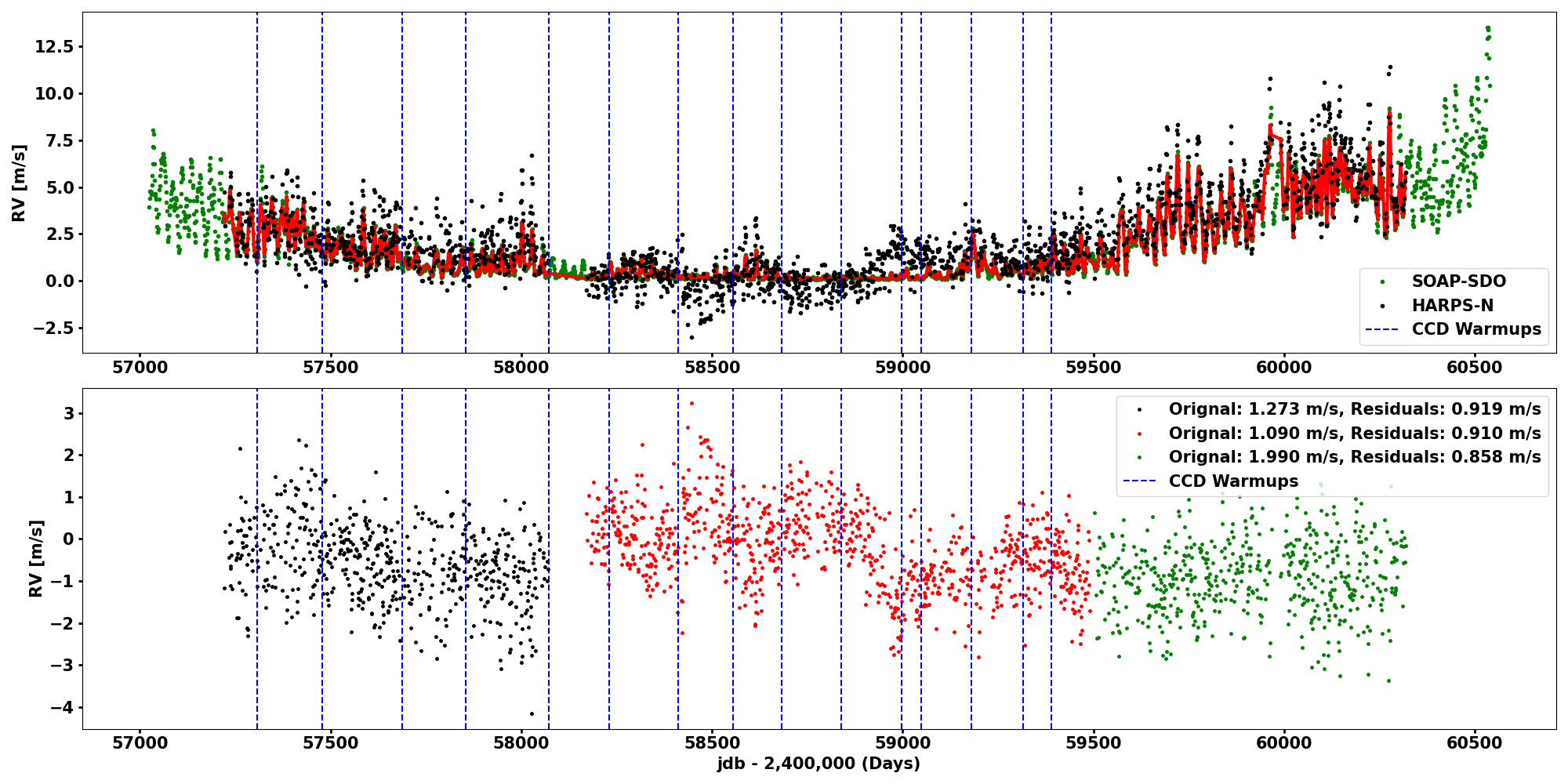}
  \caption{Same as Fig.~\ref{Fig_rv_comparison}, but simulated with the input seed spectra generated from the IAG observed solar spectra (see Sect.~\ref{sec2.2}).}
    \label{Fig_rv_comparison_IAG}%
    \end{figure}  

   \begin{figure}[htbp]
    \centering 
    \includegraphics[scale=0.3]{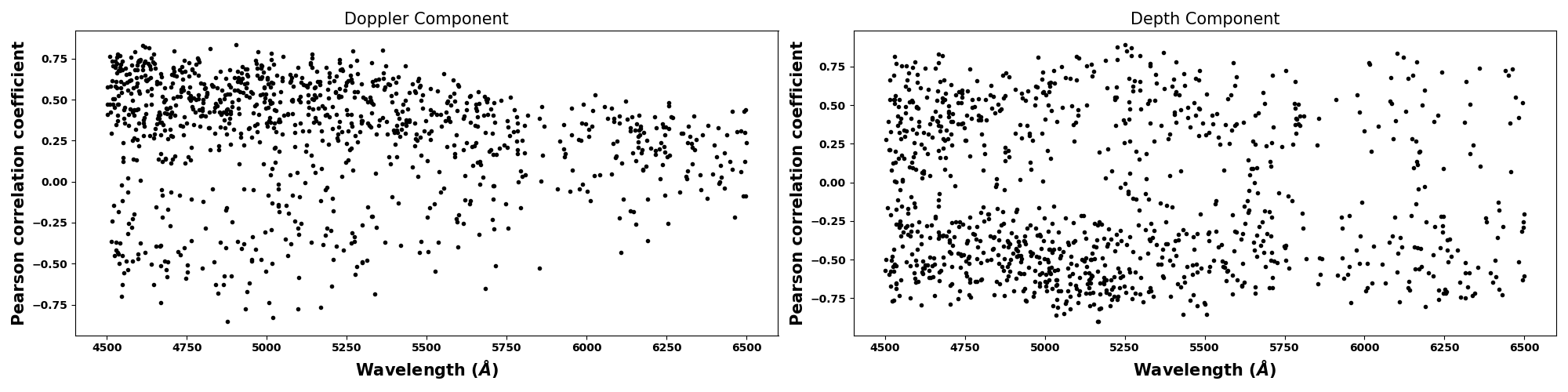}
    \caption{Same as Fig.~\ref{Fig_LBL_wave_SDO}, but simulated with the input seed spectra generated from the IAG observed solar spectra (see Sect.~\ref{sec2.2}).}
    \label{Fig_LBL_wave_SDO_IAG}%
    \end{figure}

    \begin{figure}[htbp]
    \centering 
    \includegraphics[scale=0.3]{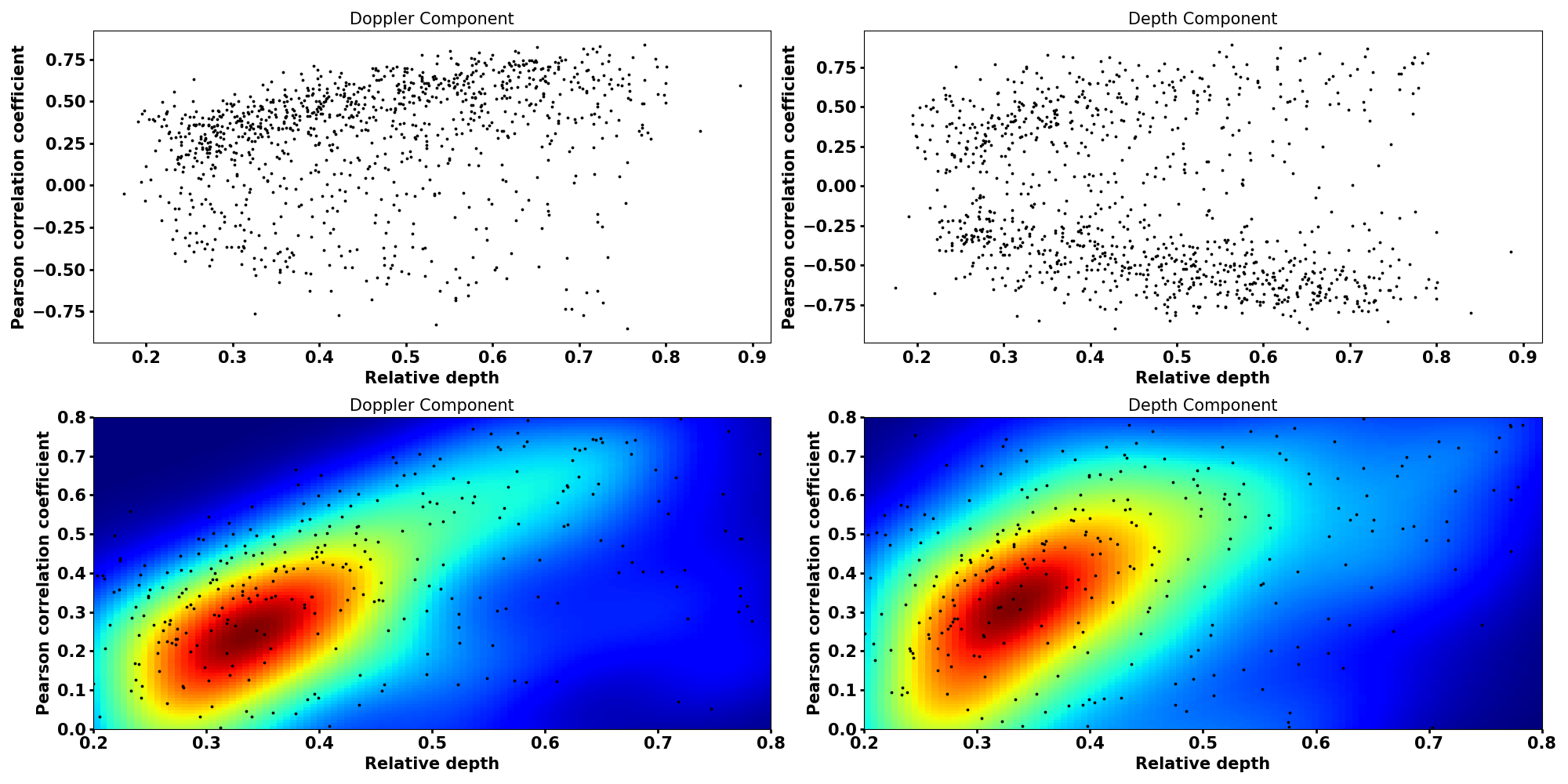}
    \caption{Same as Fig.~\ref{Fig_LBL_coeff_SDO}, but simulated with the input seed spectra generated from the IAG observed solar spectra (see Sect.~\ref{sec2.2}).}
    \label{Fig_LBL_coeff_SDO_IAG}%
    \end{figure}

\end{appendix}

\end{document}